\documentclass{emulateapj}

\def\arcmin{{$^{\prime}$}}
\def\arcsec{{$^{\prime\prime}$}}

\def\ptsec{$''\mskip-7.6mu.\,$}
\def\Msun{\,{\rm M$_{\odot}$}}
\def\Lsun{\,{\rm L$_{\odot}$}}

\newcommand{\CII}{[C {\sc ii}]}

\newcommand{\kms}{\mbox{km\,s$^{-1}$}}
\def\degr{$^{\circ}$}

\shorttitle{NGC,7538\,IRS1 - a giant N-S outflow}
\shortauthors{Sandell et al.}

\begin{document}

\title{NGC\,7538\,IRS1 - an O star driving an ionized jet and giant N-S outflow}

\author{G\"oran Sandell}
\affil{University of Hawaii, Institute for Astronomy - Hilo,  640 N. Aohoku Place, Hilo, HI 96720, USA}
\email{gsandell@hawaii.edu}

\author{Melvyn Wright}
\affil{Radio Astronomy Laboratory, University of California, Berkeley
601 Campbell Hall, Berkeley, CA 94720, U.S.A.}

\author{Rolf G\"usten}
\affil{Max Planck Institut f\"ur Radioastronomie, Auf dem H\"ugel 69, 53121 Bonn, Germany}

\author{Helmut Wiesemeyer}
     \affil{ Max Planck Institut f\"ur Radioastronomie, Auf dem H\"ugel 69, 53121 Bonn, Germany}
     
\author{Nicolas Reyes }
     \affil{ Max Planck Institut f\"ur Radioastronomie, Auf dem H\"ugel 69, 53121 Bonn, Germany}
     
\author{Bhaswati Mookerjea}
\affil{Tata Institute of Fundamental Research, Homi Bhabha Road,
Mumbai 400005, India}
          
\and     

\author{Stuartt Corder}
\affil{ALMA,  Av Apoquindo 3946 Piso 19, Las Condes, Santiago, Chile}

\begin{abstract}
NGC\,7538 IRS\,1 is a very young embedded O star  driving an ionized jet and accreting
mass with an accretion rate $>$10$^{-4}$ \Msun{}/year, which is quenching the
hypercompact \ion{H}{2} region. We use SOFIA GREAT data, Herschel PACS and
SPIRE archive data, SOFIA FORCAST archive data, Onsala 20m and CARMA
data, and JCMT archive data to determine the properties of the O star and its outflow. 
IRS\,1 appears to be a single O-star with a bolometric luminosity $>$ 1 $\times$ 10$^5$ \Lsun,
i.e.  spectral type O7 or earlier. We find that IRS\,1  drives a large molecular
outflow with the blue-shifted northern outflow lobe extending to $\sim$
280\arcsec{} or 3.6 pc from IRS\,1. Near IRS\,1 the outflow is well aligned with
the ionized jet.  The dynamical time scale of the outflow is $\sim$ 1.3 $\times$
10$^5$ yr. The total outflow mass is $\sim$ 130 \Msun. We determine a
mass outflow rate of 1.0 $\times$10$^{-3}$ \Msun{}yr$^{-1}$, roughly
consistent with the observed mass accretion rate. We observe strong high
velocity \CII\ emission in the outflow, confirming that strong UV radiation from
IRS\,1 escapes into the outflow lobes and is ionizing the gas.  Many O stars may 
form like low mass stars, but with a higher accretion rate and in a denser environment. As long
as the accretion stays high enough to quench the \ion{H}{2} region, the star will
continue to grow. When the accretion rate drops, the \ion{H}{2} region will
rapidly start to expand.
\end{abstract}


\section{Introduction}

The molecular cloud south of the \ion{H}{2} region NGC\,7538 harbors several
young massive stars, of which at least three, IRS\,1, IRS\,9 and NGC\,7538\,S,
are centers of young clusters. At a distance of 2.65 kpc
\citep{Moscadelli08} NGC\,7538 is by no means the closest high-mass star forming
region, but the extreme youth of these stars make them key targets for our
understanding of high mass star formation. NGC\,7538\,IRS\,1 was first detected
as an OH maser by \citet{Hardebeck71} indicating that there is another
\ion{H}{2}  region SE of the optically visible \ion{H}{2} region NGC\,7538. It 
was first detected in the radio at 5 GHz by \citet{Martin73}, who found three
compact radio sources at the SE edge of the compact ($\sim$ 4\arcmin{})
\ion{H}{2} region NGC\,7538. The brightest of the three, i.e., Martin's source
B, is commonly referred to as NGC\,7538\,IRS\,1, based on near/mid-IR high
spatial resolution mapping and photometry by \citep{Wynn-Williams74}. In
addition to the radio sources IRS\,1 -- 3, \citet{Qiu11} also found 8  cores at
1.3 mm using the Submillimeter Array (SMA) within a projected distance of 0.35
pc from IRS\,1, many of which are associated with H$_2$O maser emission or
outflow activity suggesting they are young pre-main-sequence stars.
\citet{Frau14} increased the number of cores to 13 by more sensitive
observations with the SMA at 880 $\mu$m. IRS\,1 is the most massive star in a
cluster with more than 150 Young Stellar Objects (YSOs) and protostars
\citep{Mallick14,Sharma17}.

The radio emission from IRS\,1 was first resolved with the Karl G. Jansky Very
Large Array (VLA) at 14.9 GHz by \citet{Campbell84}, who showed that the radio
emission consists of a compact ($\sim$0\farcs2{}) bipolar N-S core with faint
extended fan shaped lobes, suggesting an ionized outflow.  High angular
resolution observations of radio recombination lines
\citep{Gaume95,Sewilo04,Keto08} revealed extremely broad lines indicating
substantial mass motion of the ionized gas. \citet{Keto08} modeled the spectral
energy distribution (SED) of IRS\,1 as a combination of two \ion{H}{2} regions
plus dust, which could adequately fit the data available at the time. However,
\citet{Sandell09}, who analyzed high angular resolution VLA archive
data from 4.86 GHz to 43.4 GHz found that the bipolar core had a spectral index
of $\sim$ 0.8 and that the size of the core shrank as a function of frequency
with a power law index of $\sim$ - 0.8, which is what one would expect from an
ionized jet \citep[e.g.,][]{Reynolds86}, but not from an optically thick
\ion{H}{2} region. It is therefore clear that IRS\,1 drives a bipolar ionized
jet.

IRS\,1 is heavily accreting with an accretion rate of  $\sim$10$^{-3}$ - 2
10$^{-4}$ \rm M$_{\odot}$/yr \citep{Sandell09, Klaassen11, Qiu11, Beuther12,
Zhu13}, which quenches the expansion of an  \ion{H}{2} region. IRS\,1 itself is
an extremely rich maser source, with not only just the normal masers found in
high mass stars like OH, H$_2$O, and CH$_3$OH, but also rare masers like NH$_3$,
H$_2$CO, and the 23.1 GHz class II CH$_3$OH maser. The latter has to date been
only detected in three sources, see \citet{Galvan10} and references therein. 
These masers are most likely pumped by the radio emission from the jet.
IRS\,1 is almost certainly surrounded by an accretion disk, but so far there is
no firm evidence for a Keplerian accretion disk surrounding IRS\,1. The
orientation of the linear polarization of the 157 GHz CH3OH class II maser
towards IRS\,1 agrees with that of the submillimeter dust polarization,
suggesting a magnetic field oriented along the outflow axis
\citep{Wiesemeyer04}. The same holds for the 133 GHz class I maser, residing
40\arcsec\ south of IRS\,1 and possibly associated with a bow-shock in the
red-shifted outflow lobe. While it is unclear whether the 157 GHz maser
originates from an accretion disk or from the outflow. There have been many
claims that IRS\,1 is surrounded by an accretion disk in Keplerian rotation
\citep{Minier98, Minier00, Pestalozzi04, Pestalozzi09,Surcis11,Moscadelli14}
based on linear methanol maser features. \citet{Moscadelli14} identified three
maser groups and suggested that IRS\,1 is a triple system, where each high mass
YSO in the system is surrounded by an accretion disk and probably drives an
outflow. \citet{Beuther17}, who imaged IRS\,1 at 23 GHz with the VLA in the A
configuration, identified two of these maser features being close to peaks in
the free-free emission, and argued that IRS\,1 is a binary, powered by two
hypercompact \ion{H}{2} regions, each with a spectral index of 1.5 - 2. This
does not agree with the observed spectral index of IRS\,1 and with high angular
(CARMA\footnote{https://www.mmarray.org}) A-array data analyzed in this paper.
Here we will therefore re-examine whether IRS\,1 is a single star or a multiple
system.

\citet{Scoville86} using the Owens Valley Radio Observatory (OVRO) millimeter
array at 2.7 mm was the first to positively identify the high velocity CO(1-0)
emission seen toward NGC\,7538 as a bipolar outflow powered by IRS\,1 with the
outflow at a position angle (PA) of 145\degr{} (counterclockwise against north),
being red-shifted in the SE and blue-shifted to the NW with a total extent of
$\sim$ 3\arcmin, i.e., 2.3 pc. This agrees well with other studies
\citep{Fischer85, Kameya89, Davis98}. However, if this is the case, the ionized
jet and the molecular outflow are misaligned. This led \citep{Kraus06} to
interpret mid-IR speckle images as a precessing jet where IRS\,1 would undergo
non-coplanar tidal interaction with an undiscovered close companion within the
circumbinary protostellar disk. \citet{Sandell12} presented single dish James
Clerk Maxwell Telescope (JCMT) CO(3--2) data, which show blue-shifted
high-velocity gas $\sim$80\arcsec\ north of IRS\,1, which suggested that the
molecular outflow is quite large and well aligned with the ionized outflow. Here
we present an extended CO(3--2) map from JCMT, as well as \CII\ 158 $\mu$m and
high-J CO data obtained with GREAT on the Stratospheric Observatory for Infrared
Astronomy (SOFIA). Analysis of these new data sets conclusively show that IRS\,1
drives a large N-S molecular and ionized outflow.

\section{Observations}

\subsection{GREAT observations}

The NGC\,7538 IRS\,1 outflow was observed in the GREAT\footnote{GREAT is a
development by the MPI für Radioastronomie and the KOSMA / Universität zu Köln,
in cooperation with the MPI für Sonnensystemforschung and the DLR Institut für
Optische Sensorsysteme.} L1/L2 configuration \citep{Heyminck12} on SOFIA
\citep{Young12} on January 23, 2015 as part of the open time project 02\_0038.
These observations were done on a 44 minute leg at 41300 feet. The L2 mixer was
tuned to the \CII\ $^2{\rm P}_{3/2} \to\  ^2{\rm P}_{1/2}$ transition at
1.9005369 THz , while the L1 mixer was tuned to CO(11--10) at a rest frequency
of 1.26701449 THz. The main beam coupling efficiency, $\eta_{mb}$ = 0.69 for L2,
and 0.65 for L1. The Half Power Beam Width (HPBW) for L2 was measured to be
14\farcs1 for \CII\ and 21\farcs1 for CO(11-10). These observations were
complemented with observations on May 20, and 21, 2015 during the Low Frequency
Array (LFA) commissioning. The LFA array is a hexagonal array with two
polarizations co-aligned around a central pixel, providing a  2 $\times$ 7 pixel
array with the pixels separated by two beam widths.  For a more complete
description of the instrument, see \citet{Risacher16}. The LFA receiver was
tuned to  \CII, while the second channel, L1, was tuned to CO(11--10). The
measured HPBW for LFA was 15\ptsec1 and the main beam coupling efficiency,
$\eta_{mb}$ = 0.69, while the HPBW for L1 was 21\farcs1 and $\eta_{mb}$ = 0.64.
On May 20, we had a 50 minute leg at 45,000 ft. On May 21 the leg duration was
52 minutes at 43,000 ft. The precipitable water vapor was $\leq$ 9 $\mu$m on
both nights. In January 2015 (L1/L2) we did an OTF TP map with a size of
82\ptsec5 $\times$ 97\ptsec5 scanning in RA with 7\ptsec5 sampling and an
integration time of 2 second/point. The off position was at +600\arcsec,0\arcsec
relative to IRS\,1. We also took a 10 minute pointed integration on IRS\,1. In
May 2015 we did much larger maps with upGREAT. These were done in classic
on-the-fly (OTF) total power (TP) mode, with scanning in RA with a 7\arcsec\
sampling and an integration time of 2 second/point using the same off position.
The final map size in \CII\ is $\sim$ 160\arcsec\ $\times$ 240\arcsec. For
CO(11--1) the map size is 105\arcsec\ $\times$ 196\arcsec. All the data were
added together in the post processing.

The backends for both GREAT channels are the latest generation of fast Fourier
transform spectrometers (FFTS) \citep{Klein12}. These have  4 GHz bandwidth and 16384
channels, which provide a channel separation of 244.1 kHz (0.0385 km~s$^{-1}$ for
\CII{}). The data were reduced and calibrated by the GREAT team. The post
processing was done using CLASS\footnote{CLASS is part of the Grenoble Image and
Line Data Analysis Software (GILDAS), which is provided and actively developed
by IRAM, and is available at http://www.iram.fr/IRAMFR/GILDAS}. We removed
linear baselines, threw away a few damaged spectra and created maps with
rms weighting.


\begin{deluxetable*}{lllrrrrr}[b]
\tablecolumns{8}
\tablewidth{0pt} 
\tablecaption{Positions and flux densities of FORCAST mid-infrared  sources. Errors are one sigma statistical errors.\label{tbl-1}}
\tablehead{
\colhead{Source} & \colhead{$\alpha$(2000.0)} & \colhead{$\delta$(2000.0)}  & \colhead{S(7.7 $\mu$m)}& \colhead{S(19.7$\mu$m)} & \colhead{S(25.3 $\mu$m)} & \colhead{S(31.5 $\mu$m)} & \colhead{S(37.0 $\mu$m)}   \\ 
   & \colhead{[$^h$  $^m$ $^s$]}& \colhead{[$^\circ$ \arcmin\ \arcsec ]}& \colhead{[Jy]} & \colhead{[Jy]}  & \colhead{[Jy]} & \colhead{[Jy]}  & \colhead{[Jy]} 
}
\startdata
IRS\,1 & 23 13 45.37  & +61 28 10.5 & 170.7  $\pm$  3.3\phantom{0} & 240 $\pm$ 8.1\phantom{0}&  787 $\pm$ 30 & 1320 $\pm$ 28.2 & 1710 $\pm$ 50  \\
IRS\,2 &  23 13 45.44 & +61 28 20.1 & 9.4 $\pm$ 5.0\phantom{0}   & 508 $\pm$ 16.2 &  360 $\pm$ 70 & 550  $\pm$ 30.0& 540 $\pm$  90\\
IRS\,3 & 23 13 43.62 & +61 28.14.2 &  2.9 $\pm$ 0.2\phantom{0} & 85 $\pm$ 6.0\phantom{0} &  166 $\pm$ 17 & 223 $\pm$ 12.5  & 249 $\pm$ 25\\
IRS\,4 & 23 13 32.39   & +61 29 06.2 & 5.2 $\pm$ 0.3\phantom{0}  & \nodata & \nodata & 41.8  $\pm$ 0.6  & 32 $\pm$ 0.8 \\
IRS\,1E & 23 13 48.64 & +61 28 05.1 &  0.46  $\pm$ 0.10 & 1.49 $\pm$ 0.20 & 5.4 $\pm$ 0.6 & 5.3 $\pm$ 0.5 & 6.6 $\pm$ 2.1 \\
IRS\,1SE    & 23 13 48.40 & +61 27 40.3 & $<$ 0.5  & 1.11 $\pm$ 0.10&  3.3  $\pm$ 0.5 & 3.6 $\pm$ 0.5 & 1.2 $\pm$ 0.8  
 \enddata
\end{deluxetable*}

\subsection{CARMA observations}
\label{sect-CARMA}

Observations at 1.3 mm  were obtained in the A configuration on December 6, 2011
using 16 500 MHz bandwidth spectral windows, and on December 7, 2011 using 14
500 MHz windows and two 125 MHz windows which covered the H30$\alpha$
recombination line with $\sim$ 1 km~s$^{-1}$ spectral resolution. The data were 
calibrated using the calibration sources J0102$+$584, 3C\,454.3 and MWC\,349 in a
standard way using the Miriad software package \citep{Sault95}. Due to hardware
and software limitations in the 2nd and 3rd local oscillator phase tracking for
the wide band correlator during these commissioning observations in the A
configuration, the phase tracking center was offset in each spectral window.
Apart from the offset, the images in each spectral window agreed within the
noise, and we used a shift and add procedure (using the Miriad tasks IMDIFF and
AVMATHS) to average the 500 MHz spectral windows.  The December 7 2011 data were
self calibrated using the strong H30$\alpha$ emission. The December 6 2011 data
used the  continuum emission in the same spectral window in a 500 MHz bandwidth. 
The continuum images at 227.4 GHz obtained from the two data sets are consistent.

CARMA A-array observations were also obtained at 111.1 GHz on December 10, and
12, 2011, and in the B-array configuration on January 3 -- 4, and January 5 --
6,  2012. B-array observations were also obtained at  222.2 GHz on January 2 and
3, 2012. All these observations used the same backend configuration as described above. 
These data sets were reduced in Miriad following the same procedure
described in \citet{Zhu13}.

We also retrieved some mosaicked CARMA $^{12}$CO(1--0) and $^{13}$CO(1--0)
images, which allows us to get a better look of the outflow close to IRS\,1. The
observations and data reduction of these data sets are described in
\citet{Corder08}. In particular we use a $^{12}$CO(1--0) image with 4\ptsec5
angular resolution and a velocity resolution of 1.27 \kms. This image has a root
mean square sensitivity of 0.64 K per channel. The total velocity coverage of
the $^{12}$CO is from -96 \kms\ -- -16 \kms. We also retrieved a $^{13}$CO(1--0)
image made from the  C and D array configurations resulting in a  resolution of
7\ptsec9 $\times$ 7\ptsec4 PA = 55 degr. We corrected this image for missing
zero spacing by adding the $^{13}$CO(1--0) OSO map to this image using the MIRIAD
task IMMERGE. The velocity resolution of this map is 0.33 \kms\ and it covers
the velocity range from -65.8 \kms -- -45.8 \kms.

\subsection{CO(1-0) and $^{13}$CO(1-0) mapping with the 20 m Onsala Space Telescope}
\label{sect-OSO}
We observed the NGC\,7538 molecular cloud in CO(1--0) and $^{13}$CO(1--0) with
an SIS mixer on the 20 m Onsala Space telescope in Sweden from January 28 to
February 4 2010. The backend was a 1600-channel hybrid digital autocorrelator
spectrometer with 50 kHz resolution. The beam size is 33\arcsec\ and 34\arcsec\
for $^{12}$CO and $^{13}$CO, respectively. The main beam efficiency is elevation
dependent, but for the bulk of the data it was $\sim$0.42 for CO and 0.45 for
$^{13}$CO. All observations were done in Total Power position switch mode. We
mapped the cloud on a regular grid with a 15\arcmin\ spacing using an
integration time of 3 minute/position. In total we mapped an area of 240\arcsec\
$\times$ 300\arcsec\ in CO(1--0) and 400\arcsec\ $\times$ 380\arcsec\ in
$^{13}$CO(1--0). Both maps were centered on IRS\,1. The weather was generally
quite good, although some time was lost due to heavy snowfall. The system
temperature for the $^{13}$CO setting was typically between 250 - 350 K going up
to about twice that much for $^{12}$CO. All the data reduction was done in
CLASS. We removed low order baselines and occasionally standing waves. The
spectra were then coadded with RMS weighting.  The final maps were exported as
FITS cubes and imported into MIRIAD for further analysis.

\subsection{JCMT archive data}

We obtained a fully sampled CO(3--2) map from C. Fallscheer covering a 1\degr\
$\times$ 1\degr\ region of NGC\,7538. This map was observed with the Heterodyne
Array Receiver Program (HARP; \citet{Buckle09}) on the JCMT\footnote{The James Clerk Maxwell
Telescope has historically been operated by the Joint Astronomy Centre on behalf
of the Science and Technology Facilities Council of the United Kingdom, the
National Research Council of Canada and the Netherlands Organization for
Scientific Research.} at Maunakea, Hawaii.
HARP has a beamsize of 14\ptsec6 at 345.8 GHz and the main beam efficiency is
0.63. The reduction of the CO(3--2) data is described in \citet{Fallscheer13}.
The CO data cube covers a  1 GHz wide band with 2048 channels, providing a
velocity resolution of 0.42 \kms\ per channel. The 1$\sigma$ rms per channel is
$\sim$0.8 K on  the T$_{mb}$ scale. Here we only use part of the map covering
the \ion{H}{2} region and the molecular cloud southwest of it.

We also got two long integration HARP maps done in position switched jiggle mode
from Jan Wouterloot covering $\sim$ 100\arcsec\ $\times$ 100\arcsec, one in
$^{12}$CO(3--2) and one of  $^{13}$CO(3--2). Both maps have a bandwidth of 1000
MHz, i.e., a frequency resolution of 488.3 kHz, which for $^{12}$CO corresponds
to a velocity resolution of 0.42 \kms. The  $^{12}$CO map is a coadd of 22
observations, which were all of good quality, while the $^{13}$CO  is based on 6
independent observations.  The final averages were re-gridded onto a 30\arcsec\
grid. The rms noise per channel is 30  mK for CO(3--2) and 40 mK for
$^{13}$CO(3--2) respectively.
 
\begin{figure}
\includegraphics[width=8.5cm, angle=0]{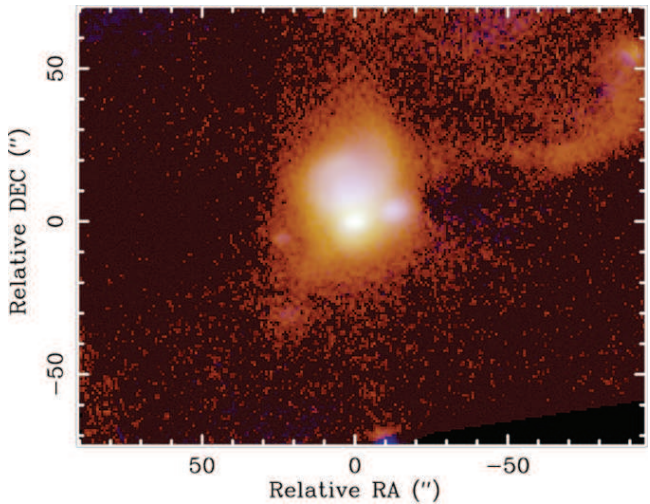}
\figcaption[]{
\label{fig-FORCAST}
Three-color image of the IRS\,1 -- 3 region generated by combining the FORCAST
19 (blue), 25 (green), and 37 $\mu$m (red) images in logarithmic scale. IRS\,1
is the strong yellow source at offset 0\arcsec, 0\arcsec. IRS\,2 at 0\ptsec7,
+9\ptsec6, is the extended blueish source with a size of $\sim$ 10\arcsec. The
third strong source, IRS\,3 is at -13\arcsec, +2\ptsec6. IRS\,4 is at the
northwestern part of the image and not fully covered in this image. IRS\,1\,E is the faint source east of IRS\,1 and IRS\,1\,SE is not really visible in this color image. }    
\end{figure}

\subsection{FORCAST archive data}

A single field around IRS\,1 was observed with the Faint Object infraRed Camera
(FORCAST) on SOFIA  on two flights in 2013, September 10 and September 13. These
observations were part of the cycle 1 project 01\_0034 (A. Tielens). We
retrieved the fully calibrated level 3 data from the SOFIA archive. The level 3
data products are pipeline reduced  and calibrated data, which require no
further processing, see \citet{Herter13} for details about data reduction and
calibration. FORCAST is a dual-channel mid-infrared camera covering 5 - 40
$\mu$m with a suite of broad and narrow-band filters \citep{Herter12}. Each
channel consists of  256 $\times$ 256 pixels. The pixel size after pipeline
processing is 0.768\arcsec. Both channels can be operated simultaneously by use
of a dichroic mirror internal to FORCAST. FORCAST is diffraction limited at
wavelengths longer than 15 $\mu$m. At 7.7 $\mu$m the beam size is 3\ptsec0, at
19.7 $\mu$m and at 37.1 $\mu$m it is 2\ptsec9 and 3\ptsec6, respectively. The
observations on September 10 were done on a 1 hour leg at 38,850 feet in dual
channel mode with filter combinations 19.7 $\mu$m/31.5 $\mu$m and 25.3
$\mu$m/37.1 $\mu$m in C2NC2 mode. Each filter combination was observed for 8
cycles with 30 second on-time each cycle. The observations on September 13 were
done on a half hour leg at 39,000 feet. These observations were done in single
channel mode with the 7.7 $\mu$m filter. Most of the leg was lost due to
instrument problems, resulting in only 3 integration cycles of 30 second on-time
each cycle. Due to field rotation and boresight acquisition, some images
partially capture IRS\,4 at the western edge of the image, while some see
IRS\,11 and South at the southern border of the images. These sources are cut
out in the coadded images, which only adds data common to each cycle. Even
though IRS\,1 and IRS\,2 were completely saturated in the {\it Spitzer} IRAC
images, the FORCAST images reveal no additional sources close to IRS\,1 and
IRS\,2. A three color image in Fig.~\ref{fig-FORCAST} shows that IRS\,1 --
IRS\,3 completely dominate the emission  and are surrounded by much fainter
diffuse extended emission. The hot PDR between the NGC\,7538 \ion{H}{2} extends
to the west and curves up to IRS\,4, only partially seen in this image. Aperture
photometry is presented in Table~\ref{tbl-1}. Because IRS\,1 overlaps with
IRS\,2, we used a small aperture and empirical aperture corrections. IRS\,2 has
a size of $\sim$ 10\arcsec, while IRS\,1 appears marginally extended. All other
sources are unresolved. The errors for IRS\,4 maybe underestimated, 
because it is at the edge of the field, and the same is true for IRS\,2, which is extended and surrounded 
by bright emission from the surrounding cloud.

\subsection{Herschel Archive data}
\label{sect-Herschel}

We have retrieved Herschel PACS and SPIRE data from the Herschel data archive.
There were two different observing projects, which covered NGC\,7538 in
PacsSpire parallel mode. F. Motte's  HOBYS Key Programme (OBS ID 1342188088 \&
1342188089) observed NGC\,7538 on December 14 2009 with slow scan speed
(20\arcsec{}/s) and the PACS blue channel set to 70 $\mu$m, i.e., PACS observed
70 and 160 $\mu$m, while SPIRE always covers all three bands; PSW (250 $\mu$m),
PMW (350 $\mu$m) \& PLW (500 $\mu$m) simultaneously.  Preliminary analysis of
these data have been presented by \citet{Fallscheer13}. The second project, S.
Molinari's OT3 project  (OBS ID 1342249088 \& 1342249089) was observed on August
5 2012. It had the same instrument setup, but used fast scanning (60\arcsec{}/s)
and covered a much larger area. However, since NGC\,7538\.IRS\,1 was severely
saturated in the HOBYS SPIRE 250 $\mu$m image, the region was re-observed as an
OT2 project (P. Andr\'e, OBS ID 1342239268) on February 13 2012. This map was
done  in small map cross scan mode with a scan rate of 30\arcsec{}/s for high
dynamic range. Here we only make use of the PACS data from Motte's program, i.e.
AORs 1342188088 \& 1342188089) and Andr\'e's high dynamic range SPIRE data.
Photometry using a combination of PSF fitting and aperture photometry is
presented in Table~\ref{tbl-2}. We also integrated the emission over the cloud
core surrounding IRS\,1 using the STARLINK application GAIA. At 70 $\mu$m the core has an equivalent radius 
of $\sim$20\arcsec, while the emission is much more extended at 160 $\mu$m, where the radius is $\sim$25\arcsec. The size of the
core is similar in the SPIRE images. The integrated PACS and SPIRE flux densities
are given in Table~\ref{tbl-2}.

\begin{deluxetable}{lcc}[h]
\tabletypesize{\scriptsize}
\tablecolumns{3}
\tablecaption{Far infrared photometry of IRS\,1 (PACS and SPIRE) and integrated flux densities of the IRS\,1 core
 derived from maps, see Section~\ref{sect-Herschel}. \label{tbl-2}}
\tablehead{
\colhead{Filter} & \colhead{Flux Density} & \colhead{Integrated flux}  \\
\colhead{[$\mu$m]}   & \colhead{[Jy]}   &   \colhead{[Jy]}        
}
\startdata
\phantom{0}70    & 6930 $\pm$ 600 & 7890 $\pm$ 790\\
160  & 3236 $\pm$ 300 & 5550 $\pm$ 500 \\
250  & 1083 $\pm$ 110 & 2760 $\pm$ 140 \\
350    &  \phantom{0}552 $\pm$ \phantom{0}55 & 1236 $\pm$ \phantom{0}60 \\
500   &  \phantom{0}190 $\pm$ \phantom{0}20 &  \phantom{0}407 $\pm$ \phantom{0}20 
\enddata

\end{deluxetable}

\section{The nature of NGC\,7538 IRS\,1}
\label{sect-nature}
\begin{figure}[h]
\centering
\begin{minipage}{\columnwidth}
\includegraphics[width=5.5cm,angle=-90]{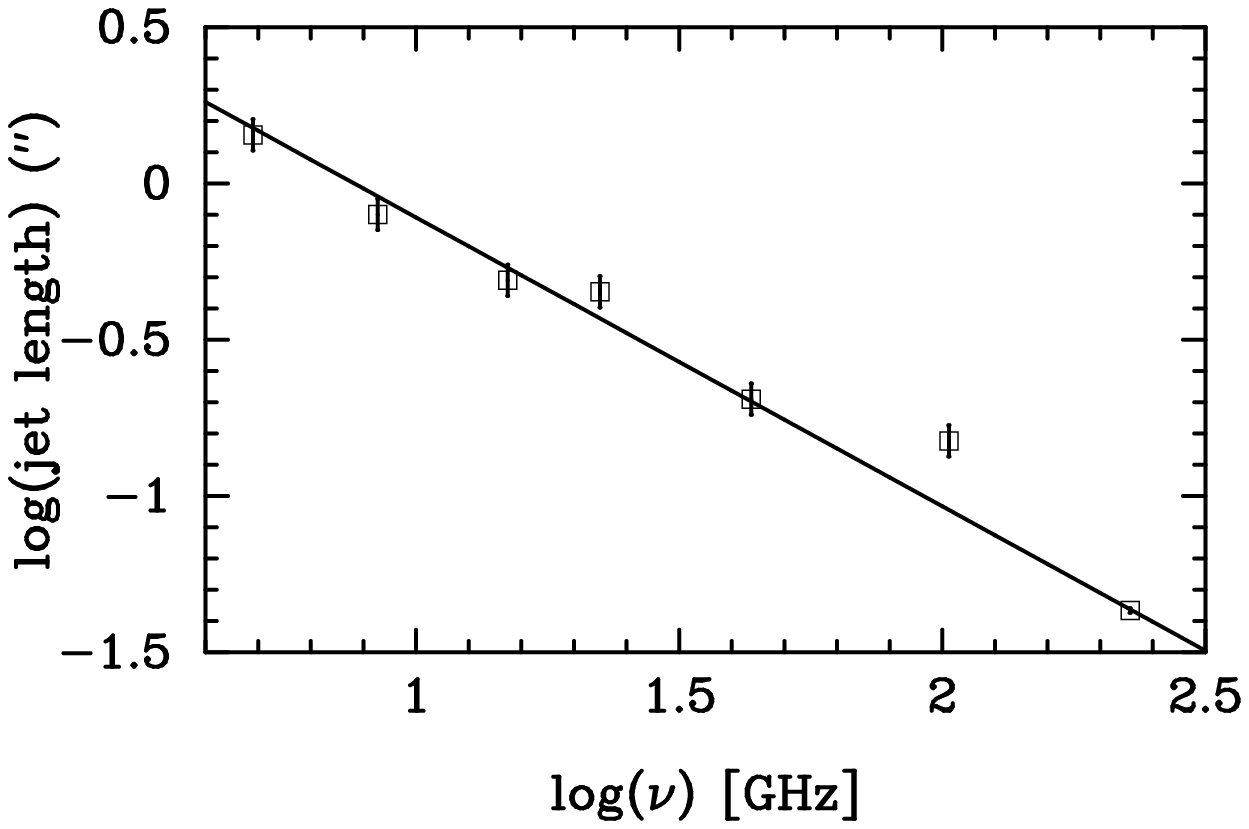}
\end{minipage}

\begin{minipage}{\columnwidth}
\includegraphics[width=5.3cm,angle=-90]{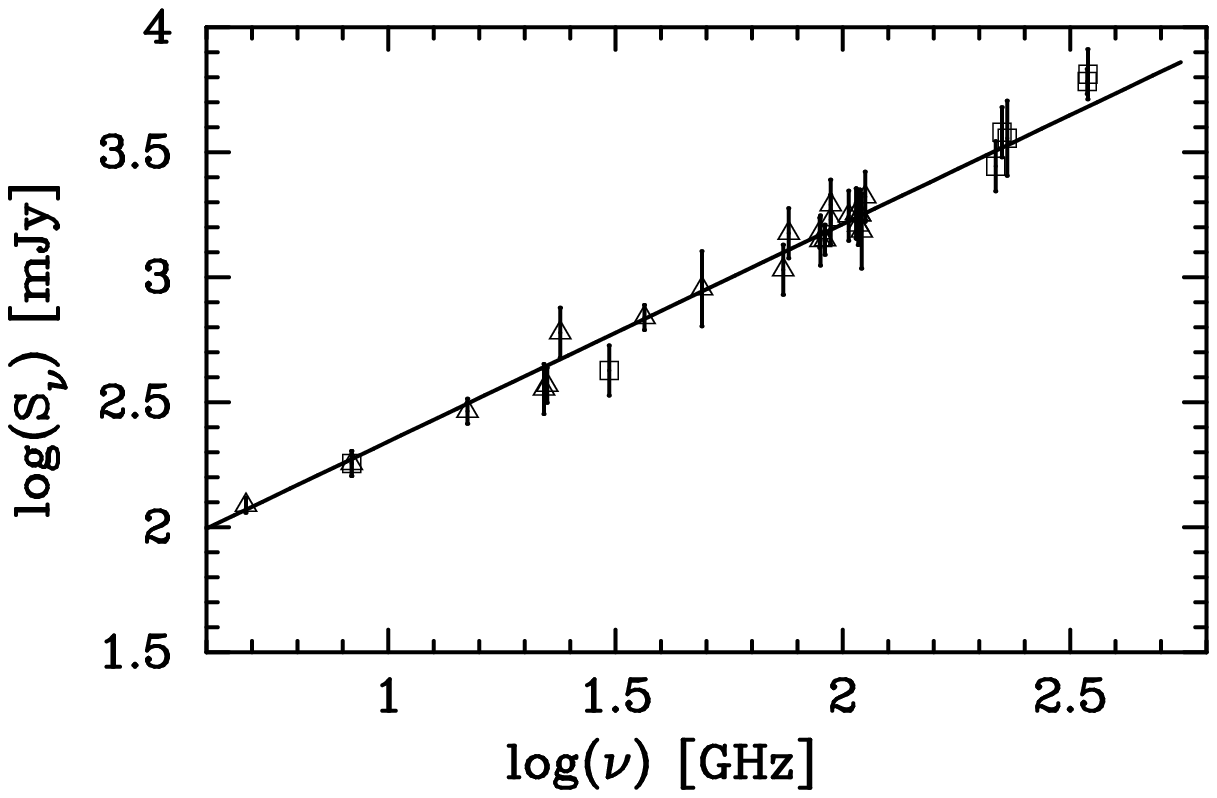}
\end{minipage}
\caption{{\it Top:}A least-squares power law fit  to the length of the compact
core surrounding IRS\,1 as a function of frequency.  The length of the core
falls off as $\nu^{-0.92 \pm 0.02}$, where $\nu$ is the frequency in GHz. 
{\it Bottom:} A least-squares power law fit to total flux densities from
published observations  of IRS\,1 from 4.8 - 115 GHz. The fit gives a spectral
index of 0.87 $\pm$ 0.03. The flux densities in the 200 and 300 GHz windows
(open squares) were excluded from the fit. Clear excess from dust emission is
only seen at frequencies above 300 GHz.
}
\label{fig-sizesed}
\end{figure}

\citet{Sandell09} proposed a model, where the free-free emission from IRS\,1 is
dominated by a collimated ionized wind driving an ionized north-south jet. This
is the only model that can explain all the observed characteristics of IRS\,1.
The model satisfactorily explains the observed morphology of IRS\,1, including
the shrinking in size of the hypercompact \ion{H}{2} region at higher
frequencies due to the outer parts of the jet becoming optically thin with
increasing frequency. It also explains the observed spectral index with
no turnover at high frequencies, as expected in a traditional \ion{H}{2} region.
Further, the extremely broad recombination lines seen towards IRS\,1
\citep{Gaume95,Keto08} and the observed variations  in flux density of time
scales of 10 years or less \citep{Franco-Hernandez04} are consistent with the
proposed model. Since there have been quite a few observations of IRS\,1 since
2009, we recomputed the spectral index by fitting the observed total flux
densities from 4.8 GHz to 115 GHz. At these frequencies the free-free emission
completely dominates and there is no contribution from dust. If there was a
contribution from dust emission at 100 GHz, it would be quite noticeable at 220
GHz, since dust emission has a spectral index of 3 -- 5. We now derive $\alpha$
= 0.87 $\pm$ 0.03, which is slightly steeper than \citet{Sandell09}
derived. In  Fig \ref{fig-sizesed} we have also included all recent flux
densities from interferometric observations at mm-wavelengths.  Our analysis
indicates that we only pick up a clear excess due to dust emission at
frequencies higher than 300 GHz. We have also redone our fit to the size
(length) of the compact core as a function of frequency including the A array data presented in this paper. We find that the size of the core
falls off as $\nu^{-0.92 \pm 0.02}$, which is consistent with a jet
\citep{Reynolds86}.  If we assume that the jet is launched from a radius of  10
-- 20 AU, this fit predicts that the free-free emission will not become
completely optically thin until somewhere between 700 -- 1500 GHz. After that
the SED would follow the typical $-$0.1 law for optically thin free-free
emission.

\begin{figure}
\includegraphics[ width=2.8cm, angle=-90]{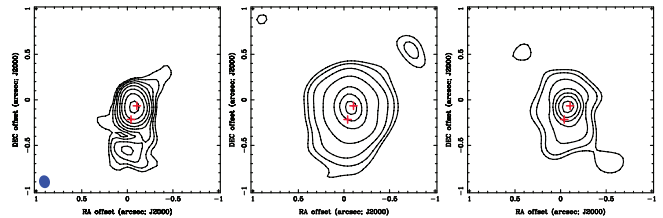}
\figcaption[]{
\label{fig-image} 
This three panel image shows the VLA 43 GHz image in the left
panel. The middle and right panel are CARMA 111.1 GHz A + B array image and a
222.2 GHz B-array image convolved to the same beam size as the VLA image
(0\farcs14 $\times$ 0\farcs12 at P.A. 14.5\degr{}). The beam is plotted in the
lower left corner of the VLA image. The red $+$ signs mark the positions of the
\citet{Beuther17} continuum sources cm1 and cm2. The position of source cm1
agrees within errors with what we identify as IRS\,1. In all these images the
reference position (0\arcsec,0\arcsec{}) is at RA (J2000) = 23$^{\rm h}$13$^{\rm
m}$45$^s.$383, Dec (J2000) = $+$61\degr\ 28\arcmin{}10\farcs506.
}
\end{figure}

\begin{figure}
\centering
\includegraphics[width=4.2cm, angle=0]{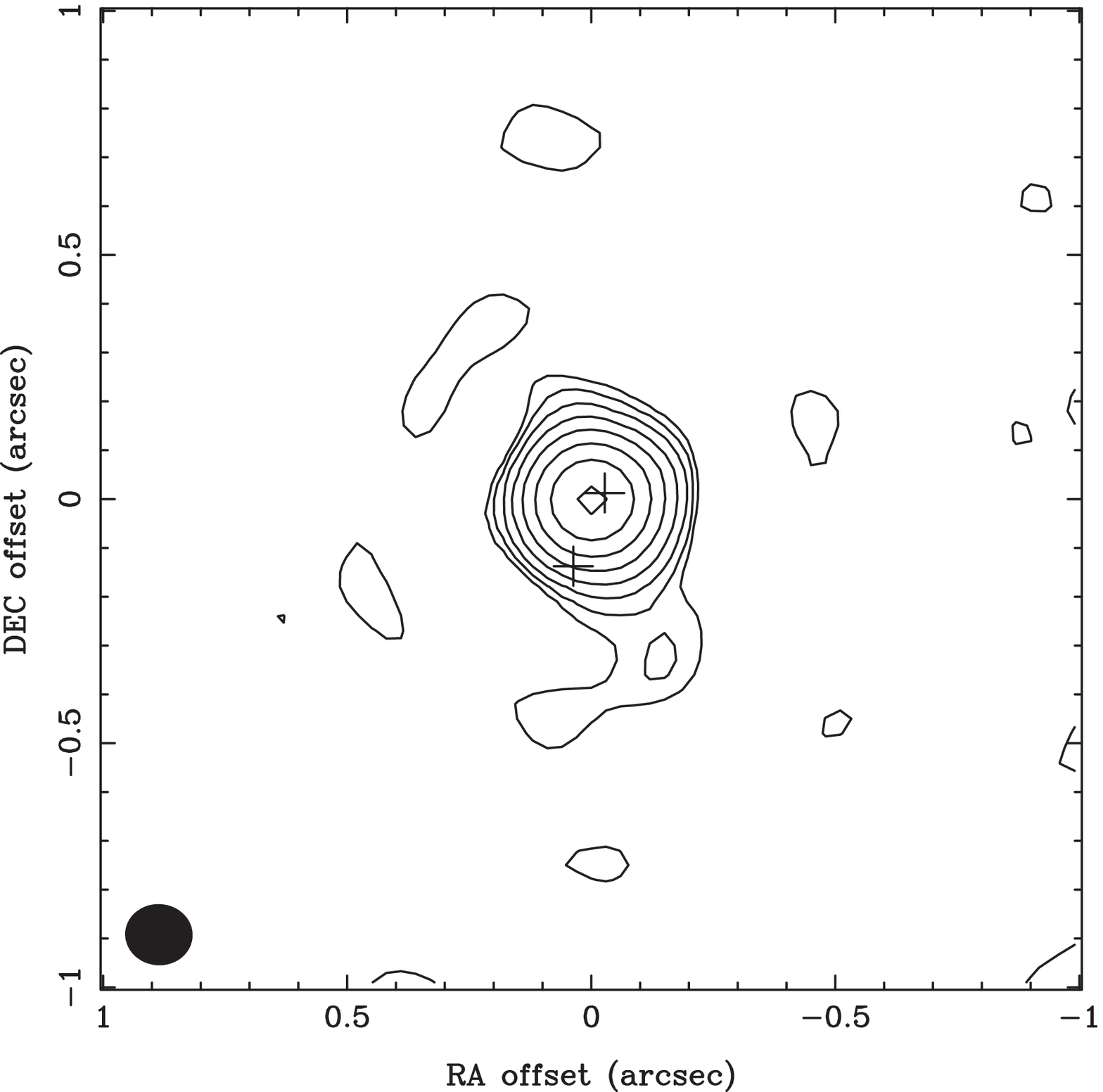}
\includegraphics[width=4.1cm, angle=0]{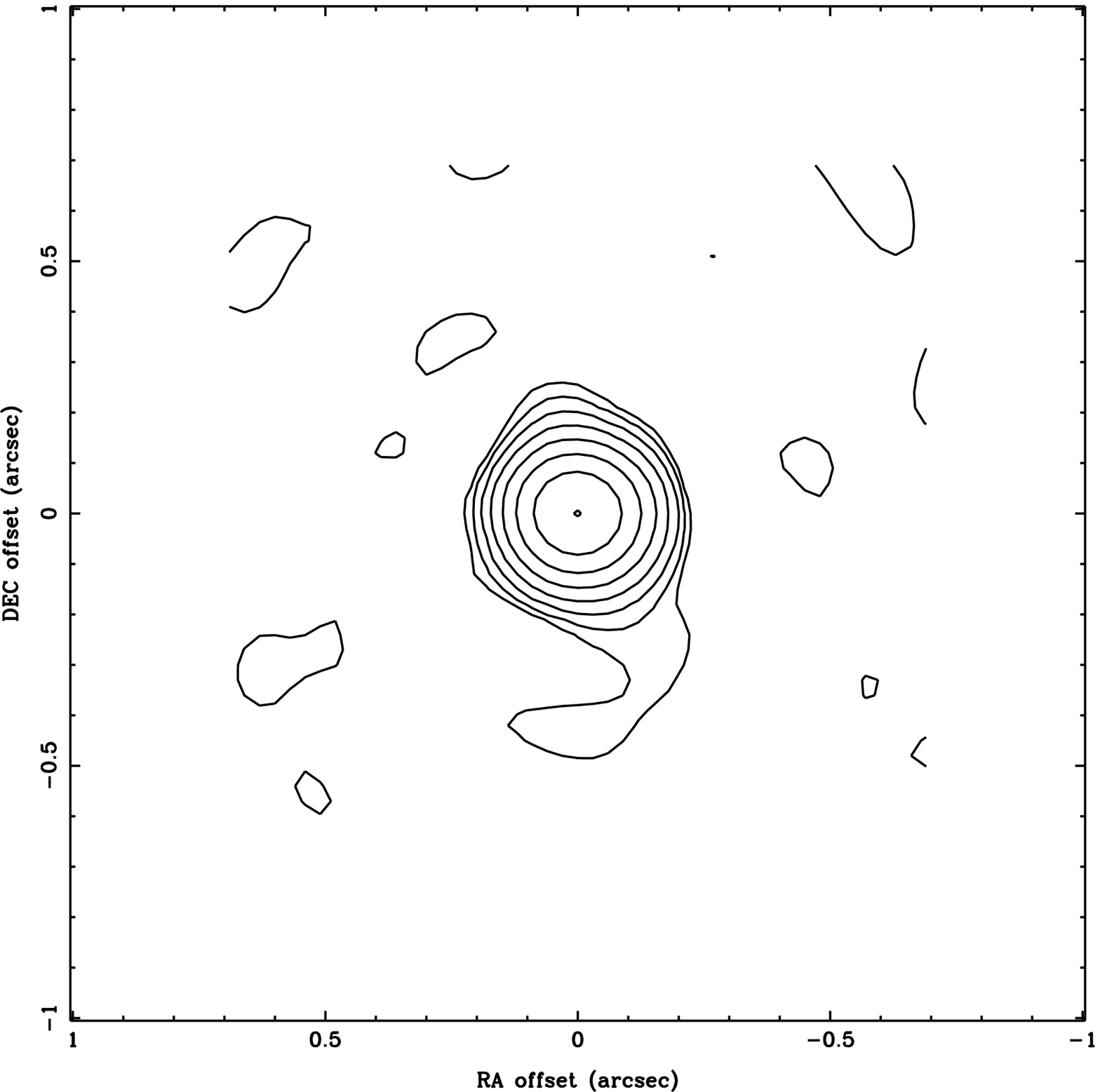}
\figcaption[]{
\label{fig-227} High spatial resolution CARMA A array wide band images at 227.4
GHz, both formed by shift and add  and reduced independently, see
Section~\ref{sect-CARMA} Both images look essentially identical. {\it Left:} The
December 7 continuum image, which used H30$\alpha$ for selfcal. The synthesized
beam is 0\farcs15 $\times$ 0\farcs13 P.A. = 89\degr. Here we also marked the
positions of the \citet{Beuther17} continuum sources cm1 and cm2. {\it Right:} The
December 6 continuum image, which used a 500 MHz wide continuum channel for
selfcal. The synthesized beam size is 0\farcs16 $\times$ 0\farcs13 P.A. =
86\degr.
}
\end{figure}

\begin{deluxetable*}{llccc}[t]
\tabletypesize{\scriptsize}
\tablecaption{Gaussian fits to the IRS\,1 core. \label{tbl-3}}
\tablehead{
\colhead{Freq}& \colhead{Array}& \colhead{uv-range}& \colhead{Synthesized Beam}& \colhead{Size} 
}
227.4 & A\tablenotemark{a} & \phantom{0}600 - 1400 k$\lambda$  & 0\farcs16 $\times$ 0\farcs13 P.A. = $+$86\degr & 0\farcs101 $\pm$  0\farcs003 $\times$ 0\farcs080 $\pm$ 0\farcs003 P.A. = \phantom{0}$+$5\degr  $\pm$ \phantom{0}6\degr \\
227.4 & A\tablenotemark{b} & \phantom{0}600 - 1400 k$\lambda$  & 0\farcs15 $\times$ 0\farcs13 P.A. = $+$89\degr & 0\farcs099 $\pm$  0\farcs004 $\times$ 0\farcs090 $\pm$ 0\farcs004 P.A. = \phantom{0}$-$6\degr  $\pm$ 13\degr \\
227.4 & A\tablenotemark{b}  & 800 - 1400 k$\lambda$  & 0\farcs1 $\times$ ~0\farcs1 ~P.A. = \phantom{-}\phantom{0}0\degr  & 0\farcs043 $\pm$ 0\farcs007  $\times$   0\farcs026 $\pm$ 0\farcs007 P.A. = $-$15\degr  $\pm$ 15\degr  \\
222.2 & B & full                                  &0\farcs29 $\times$ 0\farcs23 P.A. =  $-$85\degr & 0\farcs14\phantom{0} $\pm$ 0\farcs01\phantom{0} $\times$  0\farcs09\phantom{0}  $\pm$ 0\farcs01\phantom{0} P.A. = $+$6.1\degr $\pm$ 8\degr\\ 
111.1 & A & full                              &   0\farcs32 $\times$ 0\farcs28 P.A. =  $-$83\degr & 0\farcs15\phantom{0} $\pm$   0\farcs01\phantom{0}  $\times$  0\farcs05\phantom{0}   $\pm$   0\farcs01\phantom{0} P.A. = $+$0.9\degr $\pm$ 5\degr
\enddata
\tablenotetext{a}{Dec 5}
\tablenotetext{b}{Dec 6}
\end{deluxetable*}

Fig.~\ref{fig-image} shows the 43 GHz image from \citep{Sandell09} together with
a high resolution CARMA A + B array images at 111.1 GHz and a CARMA B-array
image at 222.2 GHz. All three images show  a compact core centered on IRS\,1
surrounded by faint extended emission to the south, i.e., what \citet{Gaume95}
called ``south spherical''. At 111.1 GHz the morphology of the emission looks
very similar to the 43 GHz VLA image, suggesting that all the emission is
free-free emission. At 222.2 GHz there is a hint of a more east-west structure
as well as a more pronounced tail to the southwest.  It is therefore likely that
we start to pick up some dust emission in the interface region between the
free-free outflow and the surrounding dense infalling envelope. It is unclear
whether any of this dust emission is due to the accretion disk, which should be
aligned east-west, i.e., orthogonal to the outflow. This faint emission
is more extended in the E-W direction and there is no sign of the northwestern
extension. In the south the emission is less extended and curves more to the
southwest. At the highest angular resolution, i.e. the CARMA A-array images at
227.4 GHz, the faint extended emission is filtered out, and we only see a
compact core with a tail towards the southwest, see Fig.~\ref{fig-227}.

To determine the size of the compact structure seen at 227.4 GHz, we created
images with three different weightings, robust = -2, 0.5 and 2, with the full
uv-range (40 -- 1400 k$\lambda$), as well as restricting the short spacings to
500, 600, and 800 k$\lambda$. The size of the IRS\,1 core shrinks in size with
improved spatial resolution, although the position angle stays the same, i.e.
roughly north south. The core is barely resolved for the uv-range 800 -- 1400
k$\lambda$, where the beam size is $\sim$ 0\farcs1, see Table~\ref{tbl-3}. At
this resolution the flux density of the compact core is 1.18 $\pm$ 0.05 Jy, or
about a third of the total flux density \citep{Zhu13,Frau14}. Table~\ref{tbl-3}
also gives the source sizes for 111 GHz and the 222 GHz B-array image. For these
we determined the deconvolved source size by fitting a double Gaussian, one for
the extended free-free jet and one for the compact core. 

\begin{figure}
\includegraphics[width=8.5cm, angle=0]{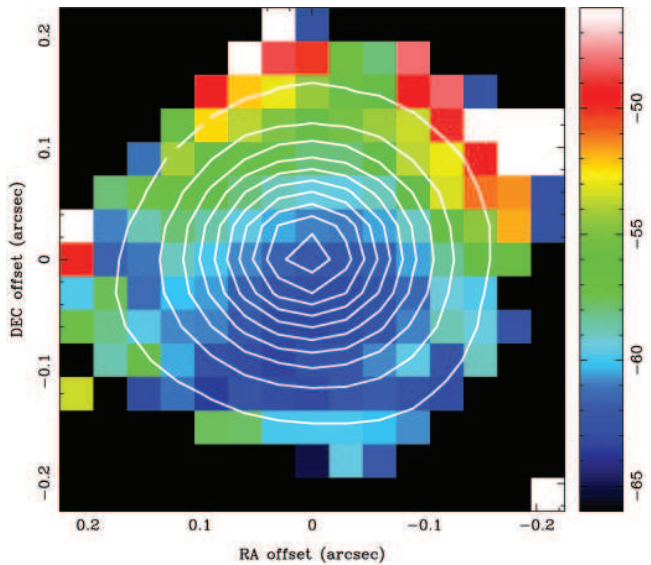}
\figcaption[]{
\label{fig-h30a_Image}
Moment 0 and moment 1 images of the the H30$\alpha$ emission made from spectral
channels above a 3-sigma threshold. The moment 0 emission is plotted with contours
on top of the moment 1 map in color. The velocity scale is shown in the
bar to the right. The images were made with a spectral resolution 
of 10 km~s$^{-1}$. A  velocity gradient in a NE direction (PA $\sim$ 30\degr{} $\pm$ 10\degr{}) 
is seen in the moment 1 image.
}
\end{figure}

We also made images of the H30$\alpha$ emission with $\sim$1 and $\sim$10
km~s$^{-1}$ spectral resolution. The H30$\alpha$ emission (line minus continuum)
is more compact than the continuum emission with a deconvolved size 72$\times$67
mas extended in a PA $-$50  $\pm$ 10 \degr. Fig.~\ref{fig-h30a_Image} shows
moment 0 and moment 1 images made from spectral channels above a 3-sigma
threshold.  A  velocity gradient in a NE direction (PA $\sim$ 30 $\pm$ 10) is
seen in the moment 1 image at both 1 and 10 km~s$^{-1}$ resolution. The gradient
was confirmed by Gaussian fits to the velocity channels. In order to check if
the observed velocity gradient might be due to some instrumental effect, we made
a spectral image for the same wide band spectral window in the December 6 2011
observations. No spectral gradient was seen in these data; Gaussian fits to the
spectral channels in the 500 MHz spectral window agrees in position within
0\farcs001. The velocity gradient of the H30$\alpha$ emission agrees quite well
in PA with the velocity gradient that \citet{Zhu13} found, 43\degr, in
subarcsecond imaging of OCS(19--18), CH$_3$CN(12--11) and $^{13}$CO(2--1).

\begin{figure}
\includegraphics[width=6.0cm, angle=-90]{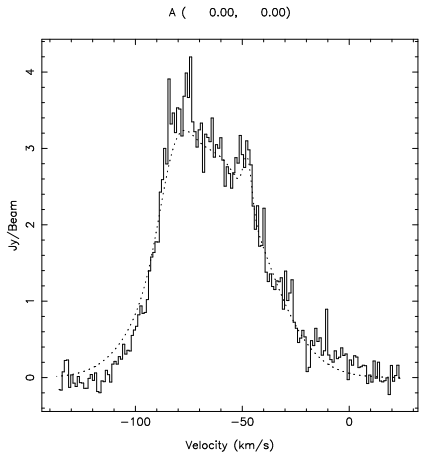}
\figcaption[]{
\label{fig-h30a_spectrum}
The H30$\alpha$ spectrum fitted with a three component Gaussian. The red and 
the blue-shifted spectral components are separated by
0\farcs005 in RA and 0\farcs009 in DEC.
}
\end{figure}

The distribution of the integrated line emission is orthogonal to the velocity 
gradient.  The structure of the H30$\alpha$ emission is unclear, because it is not 
aligned with the large scale ionized jet. The elongation and velocity gradient could arise
from a prolate rotating structure, or be due to infall onto or outflow from an
inclined disk. The fitted major and minor axis are consistent with an
inclination 21 degrees (zero = face-on), although this is not consistent with
the outflow, which suggests that the outflow is almost in the plane of the sky.
Further evidence for  an asymmetric H30$\alpha$ distribution comes from the 
H30$\alpha$ spectrum shown in Fig.~\ref{fig-h30a_spectrum}, where a three component
Gaussian fit is shown. Images of the red- and the blue-shifted spectral
components are separated by 0\farcs005 in RA and 0\farcs009 in DEC.
Gaussian fits show the red-shifted emission (-50 to 0 \kms{}) is offset to the
NE in PA 30 degrees from the blue-shifted emission (-110 to -60 \kms{}). The
red-shifted component is also somewhat larger (81 x 74 mas), than the
blue-shifted component (60 x 54 mas). The dip in the spectrum at $\sim$ -56
\kms\ might be produced by a colder envelope. Even at larger angular scales
the hydrogen recombination line emission appears rather chaotic. \citet{Gaume95}
imaged H66$\alpha$ with the VLA at 300 AU resolution. They found extremely wide
lines, 250\kms\ toward the IRS\,1 core, but no clear velocity gradient along the 
ionized jet. With the upgraded Karl Jansky VLA it would now be possible to image
recombination lines of the ionized jet further out, where
the ionized gas would not be affected by accretion and where the velocity field
might be more well ordered.

There is no sign of the three linear maser features that \citet{Goddi15} argued
to be individual \ion{H}{2} regions. Therefore, these are more likely to
be associated with the outflow \citep{DeBuizer03} or 
photoionized pre-existent clumps of molecular material \citep{DeBuizer05} rather
than self luminous objects. The 227.4 GHz observations with an angular
resolution of 0\ptsec1 show only a single compact, barely resolved source, and there is
no evidence for IRS\,1 being a binary with components cm1 and cm2 as suggested by
\citet{Beuther17}. The position of the compact core is consistent with source
cm1 within positional errors and also with the single compact source detected at
219 GHz (resolution 0\farcs33 $\times$ 0\farcs32) using NOEMA by
\citet{Beuther18}. IRS\,1 could still be a binary, but if it is, the separation
between the two stars must be less than 30 AU if the binary components have
roughly equal brightness.

IRS\,1 has a total luminosity of 1.1 -- 1.4  10$^5$ \Lsun\ (corrected to a
distance of 2.65 kpc), as determined from extinction corrected near/mid IR
observations \citep{Willner76,Hackwell82} and modeling of the observed dust
emission surrounding IRS\,1 \citep{Akabane05}. We have independently checked the
bolometric luminosity of IRS\,1 by integrating over the observed SED, which
includes accurate high angular  resolution far-infrared SPIRE, PACS, and FORCAST
images from 500 $\mu$m -- 19 $\mu$m. Using published photometry from 2.2 $\mu$
to 1.3 mm, including the photometry derived in this paper (Table~\ref{tbl-1} \&
\ref{tbl-2}), we get a bolometric luminosity of 1.0 10$^5$ \Lsun. Here we used
the PACS and SPIRE integrated flux densities in Table~\ref{tbl-2} to include
heating from IRS\,1 into the surrounding core, although it does not really make
much of a difference, because the bulk of the luminosity comes from shorter
wavelengths. This is a lower limit to the total bolometric luminosity, because
we know that a significant amount of FUV light escapes into the outflow without
heating up the cloud envelope surrounding IRS\,1.

The observed luminosity corresponds to an O7 star, if the central source is a single
star. The spectral type determined from free-free emission has resulted in
somewhat later spectral types,  O8 - B0 ZAMS \citep[see e.g.,][]{Hackwell82,
Akabane05, Zhu13}, because the ionizing flux has been estimated at a wavelength
where the free-free emission is still significantly  optically thick. Shi Hui
(2015, personal communication) derives a spectral type of $\sim$ O5.5  from
modeling high spatial resolution VLA and mm-interferometry data, which is more
consistent with the observed bolometric luminosity.

We have refined the mass estimate for the IRS1 envelope using the published
single dish data complemented with the FORCAST \& Herschel data in 
Tables~\ref{tbl-1} \& \ref{tbl-2}. Here we used a two-component graybody fit to
the observed flux densities from 1.3 mm to 20 $\mu$m, thereby avoiding
contribution from the hot dust which dominates the emission at near-IR
wavelength. For the cold extended envelope surrounding IRS\,1 our fit predicts a
size of $\sim$ 20\arcsec, a dust temperature of $\sim$ 40 K, and a dust
emissivity index, $\beta$ = 1.80, resulting in  a total mass (gas + dust) of 490
\Msun. The fit predicts a size of 5\farcs7 and a temperature of 81 K for the
warm inner envelope surrounding IRS\,1. The dust emissivity index for the warm component was
fixed at $\beta$ = 1.5. We performed a similar fit for the extended  50\arcsec\
-- 60\arcsec\ cloud core surrounding the IRS\,1 -- 3 cluster and obtained
similar temperatures and sizes for the warm dust component. The temperature is
about the same for the cold cloud core, $\sim$ 42 K, while the dust emissivity index
is slightly lower, $\beta$ = 1.6, resulting in a total mass of 1,300 \Msun.
 

\section{The IRS\,1 molecular outflow}

\begin{figure*}[]
\includegraphics[angle=0,width=17.0cm]{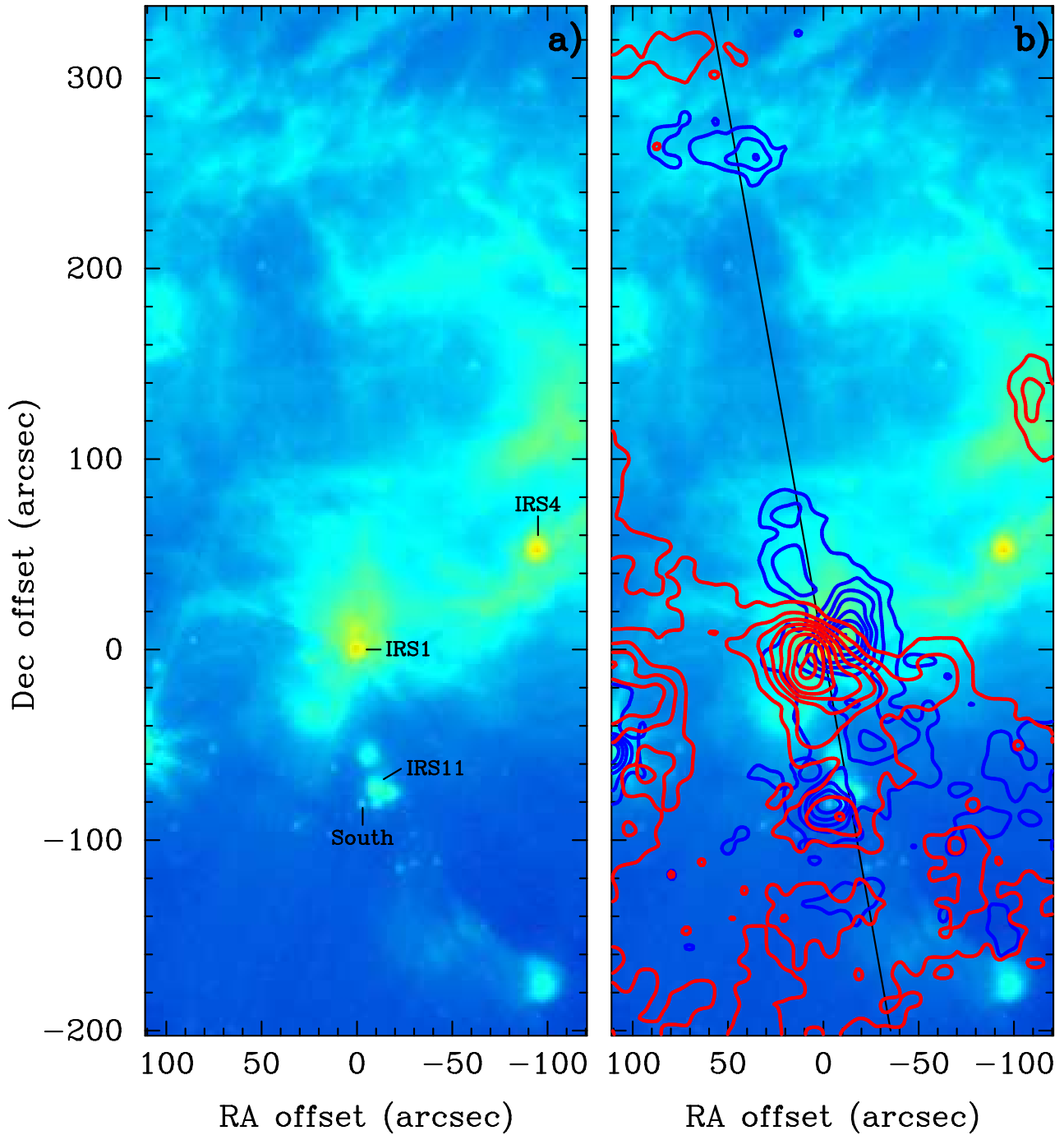}
\figcaption[]{
\label{fig-8um_co32} 
{\bf a)} 8 $\mu$m IRAC image plotted with a logarithmic stretch. IRS\,1  at
0\arcsec,0\arcsec{}  is at RA (J2000) = 23$^{\rm h}$13$^{\rm m}$45$^s$.383, Dec
(J2000) = $+$61\degr\ 28\arcmin{}10\farcs506. To the NW we see the large cavity
created by the NGC\,7538 \ion{H}{2} region, which is surrounded by PAH emission
except on the western side. There is a second cavity extending up to
250\arcsec{} from IRS\,1, which has been cleared out of the surrounding cloud
by the outflow from IRS\,1. To the southwest one can see a faint fan-shaped plume of
emission, which emerges out of the cloud and ends in a bow-shock like emission feature. 
This is almost certainly the counter-flow
to the large northern blue-shifted outflow. IRS\,9 is outside of the image to
the east, but one can see some emission from it at $\sim$ 100\arcsec,-50\arcsec.
The second bright (yellow) source is IRS\,4.  IRS\,1, IRS\,4, IRS\,11 and (NGC\,7538) South are labeled. 
{\bf b)} The same image overlaid
with high velocity CO(3--2) in contours. The blue-shifted emission is averaged 
from -82 -- -64 \kms\ and plotted with eight linearly spaced blue contours going
from 0.3 -- 3.8 K, while the red-shifted emission is averaged from -52 --
-31 \kms\ and plotted with eight contours from 0.8 -- 6.5 K. Some of the
emission seen in the red-shifted emission is due to a large extended cloud south
and southwest of IRS\,1 at -50 \kms, which overlaps with the NGC\,7538 molecular
cloud at $\sim$ -58 \kms{} (see Fig.~\ref{fig-CO32_channels}). The solid black line is drawn through IRS\,1 at a PA
of 10\degr{} and shows the approximate symmetry axis for the whole blue-shifted
outflow lobe. Close to IRS\,1 the PA is $\sim$ 0\degr\ both for the high
velocity molecular gas and the ionized jet.}
\end{figure*}

\subsection{Background}

It has been widely accepted that IRS\,1 drives a bipolar molecular outflow at a
P.A. of $\sim$ 145\degr\ \citep{Scoville86,Kameya89,Davis98,Qiu11}, although
other outflows from the young cluster surrounding IRS\,1 may also contribute
\citep{Qiu11}. \citet{Kraus06} suggested that the outflow from IRS\,1 is
precessing and has therefore created a wide-angled outflow. However, it is hard
to think of a mechanism, where the ionized outflow and the molecular outflow
would be completely misaligned.  Molecular outflows are secondary phenomena,
which are shaped by their interaction between the wind and the surrounding
cloud. It is therefore possible that the high velocity gas close to IRS\,1 may
give a distorted view of the molecular outflow, especially because of the
extreme accretion rate towards IRS\,1. \citet{Sandell12} noticed a  large fan
shaped outflow protruding out of the molecular cloud south of IRS\,1 on  {\it
Spitzer} IRAC images of NGC\,7538. This fan shaped nebula points back towards
IRS\,1.  In the north there is a large, elongated cavity, also pointing back
towards IRS\,1. The northernmost part of this cavity could not have been excavated
by the NGC\,7538 \ion{H}{2} region, because all the O stars illuminating it are
too far west of it as seen in projection, so the real separation could be even 
larger. The only star luminous enough to create such a cavity is
IRS\,1. In the 8 $\mu$m IRAC color image (Fig.~\ref{fig-8um_co32}) one can see the
cavity extending up to $\sim$ 250\arcsec, i.e., 3.2 pc,  from IRS\,1. In the
south a fan shape nebulosity emerges out of the cloud and curving
to the west. The tip of the nebulosity is quite bright and almost certainly a
bowshock, since there is no infrared source embedded in it. This bow shock, seen at
-100\arcsec, -180\arcsec\ in Fig~\ref{fig-8um_co32} is
$\sim$ 200\arcsec\ (2.6 pc) south of IRS\,1

\begin{figure}
\includegraphics[width=8.5cm, angle=0]{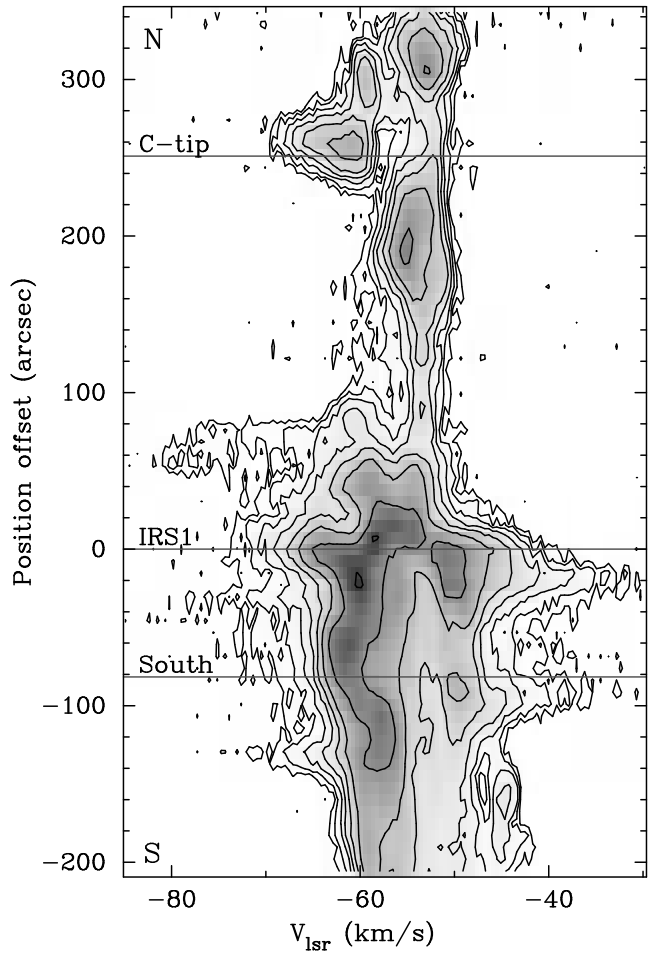}
\figcaption[]{
\label{fig-co32cut10}
Position-velocity (pv) diagram in gray scale overlaid with contours of CO(3--2).
The pv cut goes through IRS\,1 along the symmetry axis (PA = 10\degr{}) of the
IRS\,1 outflow as shown in Fig.~\ref{fig-8um_co32}\,b. IRS\,1 is at offset
0\arcsec. The tip of the blue-shifted outflow is $\sim$ 280\arcsec\ (3.6 pc)
from IRS\,1. South of IRS\,1 one can see both blue- and red-shifted emission, some of which maybe due to embedded young stars
 in the dense molecular cloud. The cut goes west of NGC\,7538\,South and misses the blue outflow lobe from the star, but picks up 
 red-shifted emission to the south of the star. We have marked and labelled the tip of the northern cavity (C-tip) seen at 8 $\mu$m (Fig. 7), 
 the location of IRS\,1, and NGC\,7538\,South with gray horizontal lines.
The contours are logarithmic with eight contours going from 0.5 -- 22 K.
}
\end{figure}

\subsection{The large N-S outflow as seen in CO(3--2)}

In Fig.~\ref{fig-8um_co32}b we have plotted  blue- and red-shifted CO(3--2)
overlaid in contours on the same 8 $\mu$m image seen in
Fig.~\ref{fig-8um_co32}\,a. The strongest high velocity emission is close to
IRS\,1 with the red (SE) and the blue-shifted (NW) emission peaks separated by
$\sim$ 30\arcsec{} at a PA of 125\degr{} - 135\degr{}, which agrees with earlier
studies. However, there is still strong blue-shifted emission extending up to
$\sim$ 80\arcsec{} north of IRS\,1, i.e., approximately where the large cavity
becomes apparent in the 8 $\mu$m image. At the northern end of this cavity
blue-shifted emission again becomes visible, where the outflow interacts with the
surrounding molecular cloud. The extent of the southern outflow is less clear in
CO(3--2). South of IRS\,1 there is both blue- and red-shifted emission,
suggesting that the dense molecular cloud has affected the outflow, making it
largely flow in the plane of the sky, i.e., perpendicular to us. We also see the
``compact'' outflow from NGC\,7538 South, which is located $\sim$ 80\arcsec{}
south of IRS\,1 and within the expected path of the southern outflow. A further
complication in the south is the rather strong emission from the -50 \kms\
cloud, see Fig~\ref{fig-CO32_channels}, which also appears to be connected to
the NGC\,7538 \ion{H}{2} region. The strong NW-SE outflow seen at low to
moderate velocities in low J CO lines is somewhat puzzling. As we will see later
on, this high velocity emission is not seen in high CO or in \CII. The most
likely explanation is that most of this outflow emission is from young embedded
low to intermediate protostars in the IRS\,1 core. Especially the northwestern blue-shifted emission
lobe appears to point toward the protostar MM\,4, as shown in the high
resolution SMA CO(2--1) image by \citet{Qiu11}.

\begin{figure}
\includegraphics[width=8.5cm, angle=0]{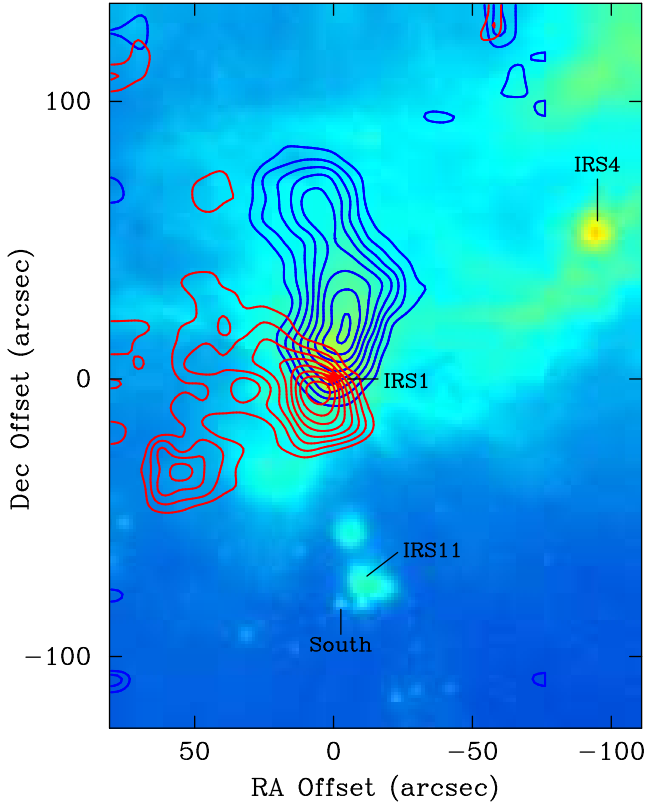}
\figcaption[]{
\label{fig-ciihigh}
High velocity \CII{} emission in blue- and red contours overlaid on a smaller
IRAC 8 $\mu$m image than in Fig.~\ref{fig-8um_co32} and plotted with a logarithmic stretch. 
IRS\,1 is at 0\arcsec, 0\arcsec\ and marked with a red star symbol. The blue-shifted emission
is averaged from -74 \kms\ - -64 \kms. and plotted with 8 linear contours
going from 0.75 K to 3.5 K. The red-shifted emission is averaged
from -52 \kms\ - -42 \kms and plotted with 8 linear contours from 0.75 K
to 2.5 K. The red-shifted peak at $\sim$  52\arcsec, -32\arcsec\} is a bright
peak in the -50 \kms\ cloud and not associated with any outflow. The red-shifted
emission NE of IRS\,1 at PA $\sim$ 45\degr{} is true high velocity emission.
IRS\,1, IRS\,4, IRS\,11 and (NGC\,7538) South are labeled.
}
\end{figure}

The position-velocity plot of CO(3--2) (Fig.~\ref{fig-co32cut10}) shows the
northern blue-shifted high velocity gas quite well. The CO emission is very
broad at the leading edge of the high velocity cloudlets at 60\arcsec{} --
80\arcsec{} north of IRS\,1, where the outflow becomes invisible, because there
is no molecular gas to interact with. The outflow becomes visible again where it
plunges into the surrounding cloud again at $\sim$ 240\arcsec{} north of IRS\,1.
Here the outflow velocity is more modest, $\sim$ 10 \kms, most likely because
there is still considerable mass in the surrounding cloud. To the south the
outflow appears both blue and red-shifted. Although the position-velocity plot
picks up the outflow from NGC\,7538\,South, this outflow cannot explain the high
velocity emission seen to the south, because the outflow from South is very
compact \citep{Sandell10}. It is therefore likely that most of this high
velocity emission is driven by IRS\,1. There is a ``bowshock'' seen in
blue-shifted emission $\sim$ 140\arcsec{} south of IRS\,1 where the outflow
emerges out of large NGC\,7538 cloud core, i.e., the dense, massive core at
$\sim$ -58 \kms.

\subsection{The IRS\,1 outflow in \CII\ 158 $\mu$m emission}
\label{sect-CII}
\begin{figure}
\includegraphics[width=8.5cm, angle=0]{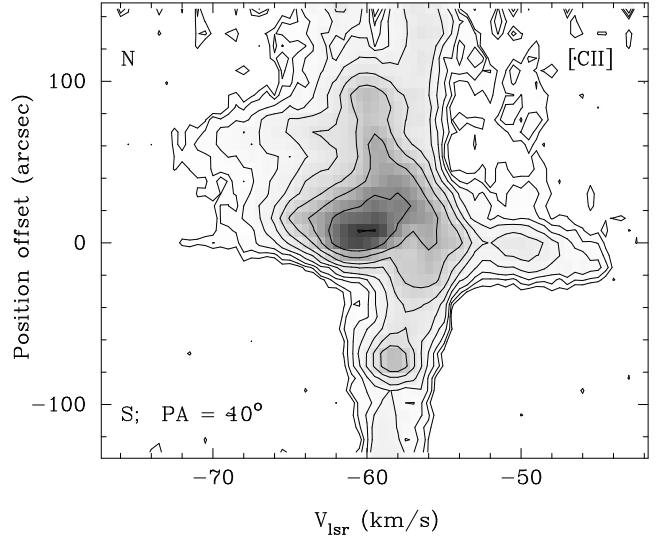}
\figcaption[]{
\label{fig-ciicut10}
Position-velocity diagram in gray scale overlaid with contours of \CII\ through
IRS\,1 at a P.A. = 10\degr{}, i.e. at the same PA as CO(3--2) in
Fig.~\ref{fig-co32cut10}. IRS\,1 is at offset 0\arcsec. The velocity structure
of the \CII{} emission looks almost identical to that of CO(3--2), except close
to IRS\,1 where one sees strong emission from the compact \ion{H}{2} region
IRS\,2, which is 8\arcsec north of IRS\,1. The contours are logarithmic with
eight contours going from 0.5 -- 22 K.
}
\end{figure}

The \CII\ fine structure line is not an outflow tracer. It is a PDR tracer
\citep{Tielens85,Hollenbach99}. Although some \CII\ emission can be produced in
shocks \citep{Flower10}, it is two to several orders of magnitude fainter than
the PDR emission. Yet \CII\ emission is seen toward many outflows
\citep{Kempen10,Podio12,Alonso17}. In most cases the \CII\ originates from PDR
emission in the surrounding cloud, but there are also some cases where one might
see \CII\ emission from the cavity walls of the outflow or from UV irradiated
shocks in the outflow\citep{Visser12,Green13,Alonso17}. Fig.~\ref{fig-ciihigh}
shows that the blue-shifted high velocity \CII\ emission fills the outflow lobe
and  looks similar to the blue-shifted CO(3--2) at high velocities. We
definitely see \CII\ emission from IRS\,2, because it is a bright compact
\ion{H}{2} region and the peak of the \CII\ emission is approximately centered
on IRS\,2.  IRS\,2, however, does not have an outflow. Observations of the [Ne
{\sc ii}] fine structure line at 12.8 $\mu$m, shows that it has ionized gas in
the velocity range -82 \kms -- -51 \kms\ \citep{Zhu08}. There is no sign of
low-velocity [C II] tracing the cavity walls of the outflow. It appears that the
strong FUV radiation from IRS\,1 ionizes some of  carbon in the fast moving CO
cloudlets resulting in strong PDR emission tracing the surface layers of the
cloudlets. The emission knot at 53\arcsec, -32\arcsec\ seen in red-shifted \CII\ 
(Fig.~\ref{fig-ciihigh}) is not an outflow. The \CII\ channel maps
 (Fig.~\ref{fig-CII_channels}) show that it is only seen in the channels centered
at -51 and -47.5 \kms. It is unclear what star illuminates  this PDR. There is not
much high velocity red-shifted \CII\ emission south of IRS\,1, although we see
strong \CII\ emission in the velocity range from $\sim$ -60 to -50 \kms\ south
and north-west of IRS\,1 (see the channel maps in Fig.~\ref{fig-CII_channels}).
Most of this emission is likely to be PDR emission illuminated by the NGC\,7538
\ion{H}{2}. There is no \CII\ emission associated with the compact
NGC\,7538\,South outflow, which is very prominent in CO(11--10), see Section
\ref{sect-CO11-10}. However, in the the channel maps we see a compact, $\sim$
10\arcsec\ knot of emission in the channel centered on -58 \kms\ about
70\arcsec{} south of IRS\,1 (Fig.~\ref{fig-CII_channels}). The position of this
knot is within 1\arcsec\ of IRS\,11, a well known young Herbig Be star near
NGC\,7538\,South. The deeply embedded IRS\,11 apparently illuminates a small
nebula, which we see in \CII. This PDR emission is also seen in the \CII\ position 
velocity diagram at $\sim$ 70\arcsec\ south of IRS\,1 (Fig.~\ref{fig-ciicut10}).

\begin{figure}[h]
\includegraphics[angle=0,width=8.0cm,angle=0]{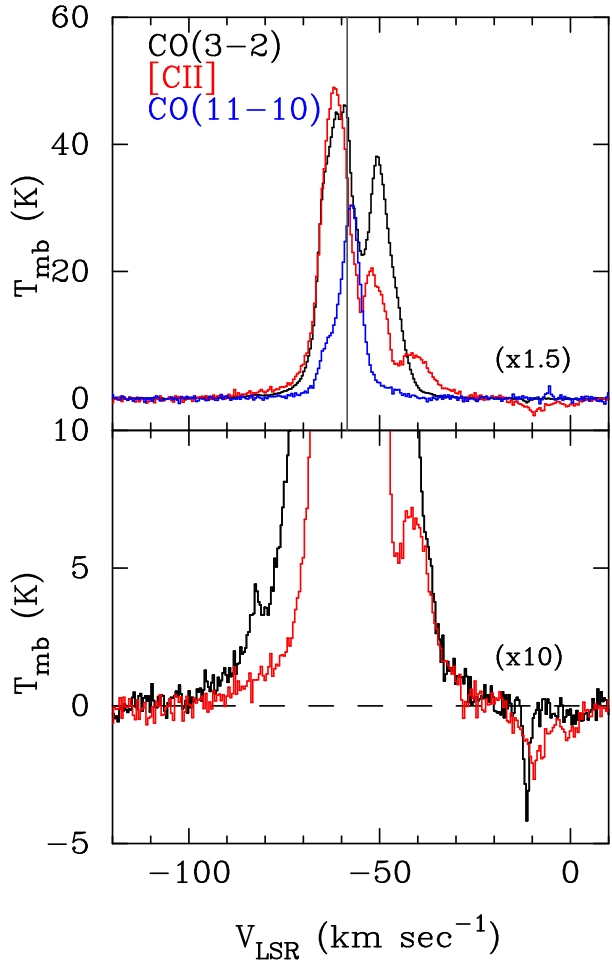}
\figcaption[]{Long integration \CII\ (red), CO(11--10) (blue), and CO(3--2)
(black) spectra in native angular resolution toward NGC\,7538\,IRS\,1. The gray
vertical line in the top panel marks the systemic velocity, -58.5 \kms. In the
top panel the CO(3-2) brightness temperature is multiplied by a factor of 1.5,
in the bottom one, which shows the high velocity wings it is scaled by a factor
of 10. The faint absorption between -12  - 0 \kms\ is due to emission or
absorption from gas in the local arm. In \CII\ it is absorption against the
strong continuum from IRS\,1. In CO(3--2) it is due to emission in the reference
position.
\label{fig-CO32CIIwings}
}  
\end{figure}

The blue-shifted emission north of IRS\,1 has the same velocity structure as
CO(3--2) (Fig.~\ref{fig-ciicut10}) with the leading bowshock at 80\arcsec\ north
of IRS1 having the highest velocities. The highest velocity seen in \CII\ is -74
\kms\ whereas it is $\sim$ -85 \kms\ in CO(3--2). The difference is most likely
due to the lower sensitivity (shorter integration times) in \CII. Using
observations with longer integration times we find that the high velocity
emission has very much the same extent in both \CII\ and CO(3--2), toward IRS,1
(Fig.~\ref{fig-CO32CIIwings}). Both \CII\ and CO(3--2) show high velocity
wings of $\pm$ 40 \kms\ or more.

There are also some clear differences between \CII\ and CO(3--2). The dominant
red-shifted (SE) and blue-shifted (NW) outflow lobes near IRS\,1 are far less
obvious in \CII\ than in CO(3--2), see Fig.~\ref{fig-co32cut10} \&
\ref{fig-ciicut10}.

\subsection{CO(11--10)}
\label{sect-CO11-10}

\begin{figure}
\includegraphics[width=8.5cm, angle=0]{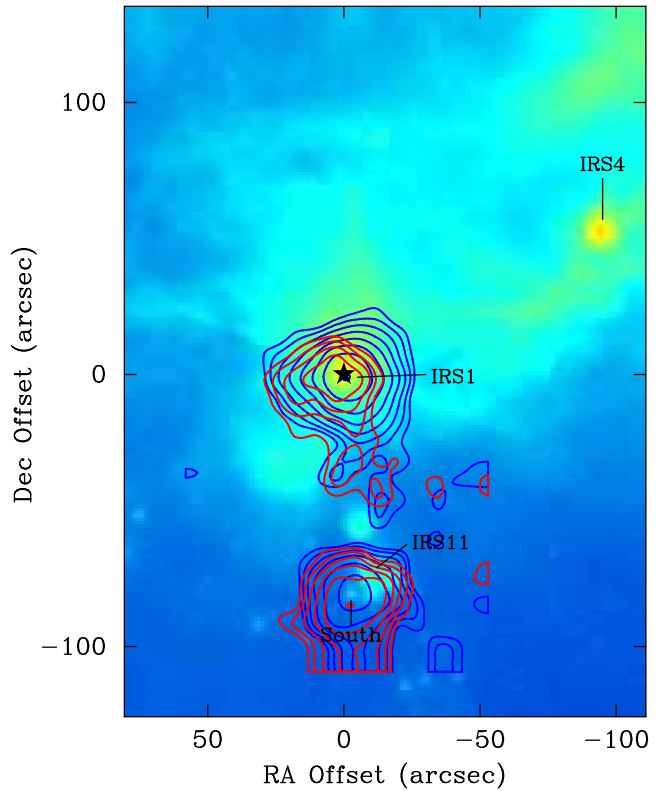}
\figcaption[]{
\label{fig-CO11_8um}
High velocity CO(11-10) emission in blue- and red contours overlaid on an IRAC 8
$\mu$m image plotted with {\bf a logarithmic stretch}. IRS\,1 is at 0\arcsec,
0\arcsec{}. The blue-shifted emission is averaged from -66 \kms\ - -60.5 \kms.
and plotted with 8 logarithmic contours going from 0.45 K to 4.8 K.
The red-shifted emission is averaged from -56 \kms\ - -50.5 \kms\ and plotted
with 6 linear contours from 0.5 K to 2.7 K. 
IRS\,1, IRS\,4, IRS\,11 and (NGC\,7538) South are labeled. }
\end{figure}

\begin{figure}
\includegraphics[width=8.5cm, angle=0]{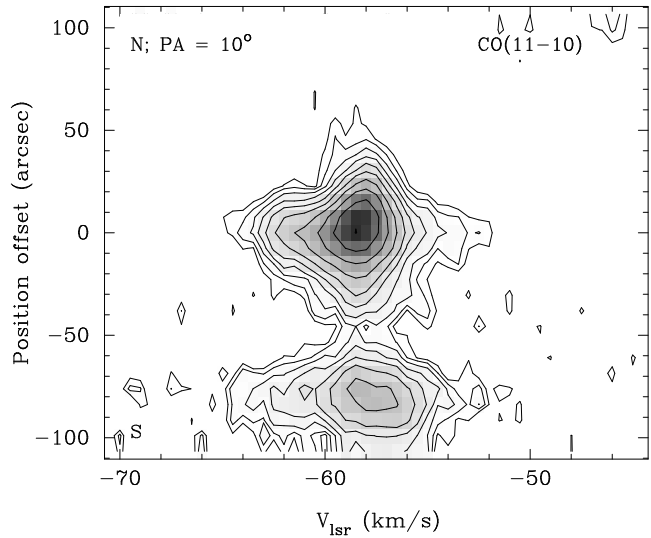}
\figcaption[]{
\label{fig-CO11cut10}
Position-velocity diagram in gray scale overlaid with contours of CO(11--10)
through IRS\,1 at PA = 10\degr, i.e. the same as for CO(3--2) and
\CII. IRS\,1 is at offset 0\arcsec. The compact outflow from the young high mass star
NGC\,7538\,South at $\sim$ 81\arcsec{} to the south is quite prominent in
blue-shifted emission. CO(11--10) emission still continues to the south. The
contours are logarithmic with ten contours going from 0.8 -- 29 K.
}
\end{figure}

CO(11--10) was observed simultaneously with \CII\ in the L1 channel of GREAT.
Since L1 is a single pixel receiver, the CO(11--10) map is not as extended as
the \CII\ map. One would expect that the emission from CO(11--10), which has a
critical density of 4 $\times$ 10$^5$ cm$^{-3}$ for a gas temperature of 100 K
\citep{Yang10}, would be quite compact. Yet CO(11--10) emission extends almost all the
way to NGC\,7538\,South, 80\arcsec\ south of IRS\,1 (Fig.~\ref{fig-CO11_8um}).
South of IRS\,1 the outflow velocities are rather modest, with line wings $\sim$
3 \kms\ in both blue- and red-shifted gas to $\sim$ 40\arcsec\ south of IRS\,1
(Fig.~\ref{fig-CO11_8um} \& \ref{fig-CO11cut10}). Further south the line
is very narrow until the emission runs into the outflow from
NGC\,7538\,South, where one sees strong blue-  and red-shifted emission. The southern
outflow lobe appears to behave very much the same way in CO(11--10) as it does
in CO(3--2) and \CII. The outflow is at near-cloud velocities or perhaps even
slightly blue-shifted, suggesting that the southern outflow lobe is close to the
plane of the sky. The emission in CO(11--10) is undoubtedly excited by the
outflow, because CO(11--10) is only seen along the expected path of the outflow.
Only hot gas in the outflow can excite CO to such high energy levels. The
velocity of the northern outflow is low and only seen to $\sim$ 40\arcsec\ north of
IRS\,1. The high-velocity cloudlets, which stand out prominently in CO(3--2) and
\CII, are too diffuse to excite CO(11--10). There is no sign of the SE-NW
outflow lobes, which dominate the outflow near IRS\,1 in low J CO lines and
other low excitation tracers, like HCO$^+$.

\subsection{The morphology of the outflow near IRS\,1; CARMA CO(1-0) and
$^{13}$CO(1-0) imaging}

\begin{figure*}
\includegraphics[width=\textwidth, angle=0]{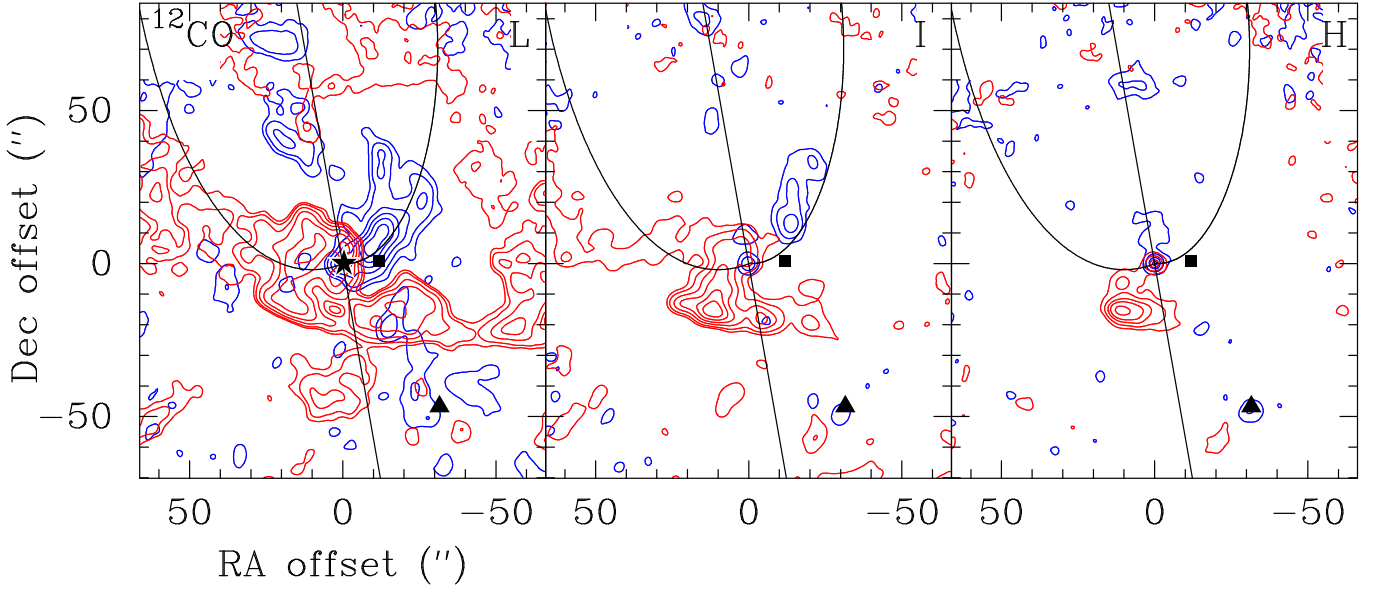}
\figcaption[]{
\label{fig-12CO_CARMA}
Contour plots of the low, intermediate and high velocity blue- and red-shifted CO(1--0) emission from
CARMA observations with 4\ptsec5 spatial resolution. Due to the primary beam 
correction the maps are noisier in the
northernmost part of the map, because it is close the HPBW of the CARMA primary
beam. The L panel shows low velocity emission, $\pm$9 \kms\ from the
systemic velocity, -58.5 \kms, averaged over a 6.4 \kms\ wide velocity interval.
The I panel  shows the outflow at intermediate velocities, $\pm$15
\kms, also averaged over a 6.4 \kms\ wide velocity interval, while the H panel
shows high velocity emission, $\pm$25 \kms\ from the systemic velocity averaged
over a 14 \kms\ wide window. At low and intermediate velocities we see
blue-shifted emission from the cloudlets in the blue-shifted outflow lobes. At
higher velocities the outflow is more jet-like and aligned with the ionized jet.
The contour levels for the low and intermediate velocities are at 0.9, 3, 7, 12,
18, and 25 K, and at 0.5, 1.5, 3, 4.5, and 7 K for the high velocity panel
(right). IRS\,1 is at 0\arcsec, 0\arcsec{} and marked with a star symbol on the
L panel. Also marked with a black square symbol is the sub-millimeter and H$_2$O
maser source MM\,4 \citep{Qiu11}, which almost certainly contributes to the
blue-shifted high velocity emission NW of IRS\,1. We also marked another H$_2$O
maser with a black triangle at offset -31\arcsec,-47\arcsec. This source appears
to be associated with a compact blue-shifted high velocity knot. We also drawn
the symmetry axis of the outflow and our best guess of the boundary between the
outflow cavity and the surrounding cloud. These are plotted in black on each
panel. 
}
\end{figure*}

\begin{figure*}
\includegraphics[width=\textwidth, angle=0]{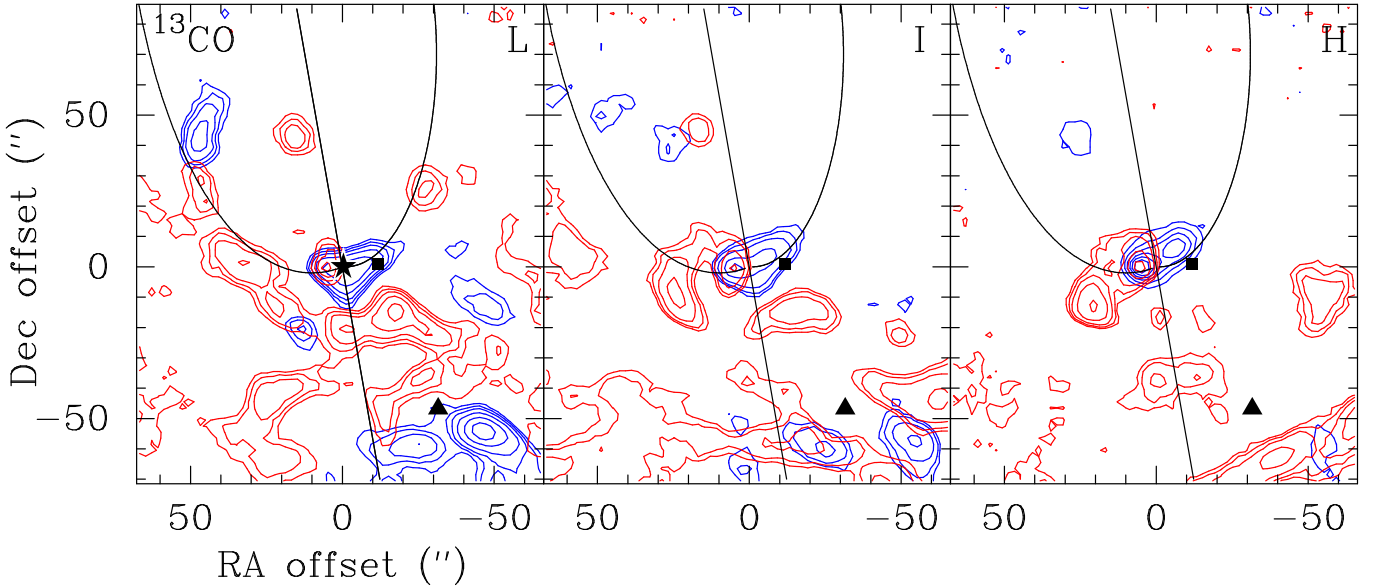}
\figcaption[]{
\label{fig-13CO_CARMA}
Contour plots of the  low, intermediate and high velocity blue- and red-shifted $^{13}$CO(1--0) emission from CARMA
observations imaged with a beam size of 7\ptsec9 $\times$7\ptsec4 p.a. = 55
\degr{} resolution and zero spacings from OSO 20 m observations plotted the same
way as in Fig.~\ref{fig-12CO_CARMA}. In the L panel the blue-shifted emission is
centered at -61.3 \kms\ and the red-shifted emission at -52 \kms, both
averaged over a 1.66 \kms\ wide window. In the I panel the center velocities
for the blue- and red-shifted emission is -63 \kms\ and -50.2 \kms with 1.66
\kms and 2.0 \kms\ wide windows for blue- and red-shifted emission respectively.
In the H panel the blue- and red-shifted center velocities are -64.8 \kms\ and
-47.5 \kms, averaged over 2 and 3.3 \kms, respectively. The contour levels 
for the low velocities are at 0.9, 1.4, 2, 3, 4.5, and 6 K, for intermediate 
velocities at 0.5, 0.9, 1.7, 3.5, and 5 K and for the high velocities (right panel)
at 0.3, 0.5, 0.8, 1.4, and 2.5 K. All the rest is plotted as in Fig.~\ref{fig-12CO_CARMA}.
}
\end{figure*}

Single dish images do not have enough spatial resolution to show how the outflow
behaves near IRS\,1. We therefore also make use of some CARMA images of CO(1--0)
and $^{13}$CO(1--0), the latter filled in for missing zero spacing with maps
from the OSO 20 m telescope, see Sect.~\ref{sect-OSO}. The emission from the
surrounding molecular cloud is very strong and it is difficult to study the
outflow at low velocities, especially in $^{12}$CO. In Fig.~\ref{fig-12CO_CARMA}
the lowest blue- and red-shifted emission starts at $\sim$6 \kms\ from the
systemic velocity of the cloud. Some of the red-shifted emission, however, is
not from the IRS\,1 outflow, but from gas associated with the expanding
\ion{H}{2} NGC\,7538 NW of IRS\,1, see e.g. \citet{Davis98}, or from dense
condensations in the ``-50 \kms{}'' cloud. The $^{13}$CO, which is more
optically thin, can probe the outflow at lower velocities but it is difficult to
evaluate, especially for red-shifted velocities, because of contamination from
the extended ``-50 \kms{}''  cloud. This cloud has several velocity components
between -52 and -45 \kms, some of which also appear to trace the PDR layer
associated with NGC\,7538. The cluster of sub-millimeter sources \citep{Qiu11}
most likely power molecular outflows as well. Some of the blue-shifted emission
NW of IRS\,1 is likely from MM\,4, which is shown as a filled black square in
Figs.~\ref{fig-12CO_CARMA} \& \ref{fig-13CO_CARMA}. Some of the red-shifted
emission south of IRS\,1 is almost certainly ``contaminated'' by emission from
one or several of these sub-millimeter sources as well.

At low velocities the $^{13}$CO emission is red-shifted East and EW of IRS\,1.
There is also some red-shifted emission to the NW, Fig.\ref{fig-13CO_CARMA}.
One can also see some red-shifted emission associated with the high velocity
cloudlets near the symmetry axis of the blue-shifted outflow. $^{12}$CO(1--0)
behaves the same way as $^{13}$CO at low and intermediate velocities, i.e. it
strongly suggests that the outflow is rotating, possibly in a spiral pattern
\citep{Wright14}. What is noticeable is that at high spatial resolution the high
velocity jet or high velocity cloudlets inside the blue-shifted outflow lobe
break up into compact clumps. At intermediate velocities the clumps
become more centered in the middle of the outflow lobe, see panel I in
Fig.~\ref{fig-12CO_CARMA}. At high velocities the blue-shifted outflow is well collimated,
almost N-S, like the free-free jet, and extending to $\sim$15\arcsec\ N of IRS\,1. 
At these velocities we no longer see the northwestern
blue-shifted outflow feature, which dominates at lower velocities. 

\subsection{What we learned about the outflow from CO(3--2), 
\CII\ and CO(11--10)}

Here we summarize what we learned about the morphology of the large scale IRS\,1
outflow from examination of CO(3--2), \CII\ and CO(11--10). The large CO(3--2)
map confirms that the large cavity north of IRS\,1 seen in IRAC images has been
created by the molecular outflow from IRS\,1. We can trace the outflow in high
velocity blue-shifted emission up to about 80\arcsec\ north IRS\,1 after which
it becomes invisible due to the low gas density of the surrounding cloud. At
the tip of the outflow, $\sim$ 250\arcsec\ to the north, the outflow again
becomes visible as blue-shifted CO(3--2) high velocity emission where it hits the dense
surrounding molecular cloud. The southern outflow from IRS\,1 is probably
deflected by the dense molecular cloud south of IRS\,1 and is almost in the
plane of the sky. Therefore it is seen as mostly blue- and red-shifted low
velocity emission. The southern outflow passes close to NGC\,7538\,South,
another young high mass star and becomes invisible when the density of the
surrounding cloud becomes to low to excite CO emission. The plume seen in IRAC
images where the outflow emerges out of the cloud appears to contain no or very
little molecular gas.

These findings are strongly supported by the smaller maps we have obtained in
\CII\ and CO(11--10). The high velocity clouds or cloudlets in the northern
outflow lobe are also seen in \CII, where they show the same morphology and
velocity structure as seen in the CO(3--2) high velocity emission. This suggest
that they are illuminated by the strong UV radiation escaping from IRS\,1 into
the outflow. They are not seen in CO(11--10), most likely because they have too
low gas density to excite CO(11--10). The outflow from IRS\,1 also has modest
blue- and red-shifted velocities south of IRS\,1 in \CII\ and CO(11--10)
confirming that the southern outflow lobe is almost in the plane of the sky.
However, since we trace CO(11--10) emission  almost all the way to NGC\,7538\,South,
this emission must be excited by IRS\,1, since CO(11--10) has a critical density
of 4 $\times$ 10$^5$ cm$^{-3}$ and an upper energy level of 304.2 K. No other
mechanism can produce such warm high density gas in this region, except
radiation from IRS\,1.

\subsection{Outflow properties}
\label{sect-Outflow_prop}

\begin{figure}
\includegraphics[width=8.0cm, angle=0]{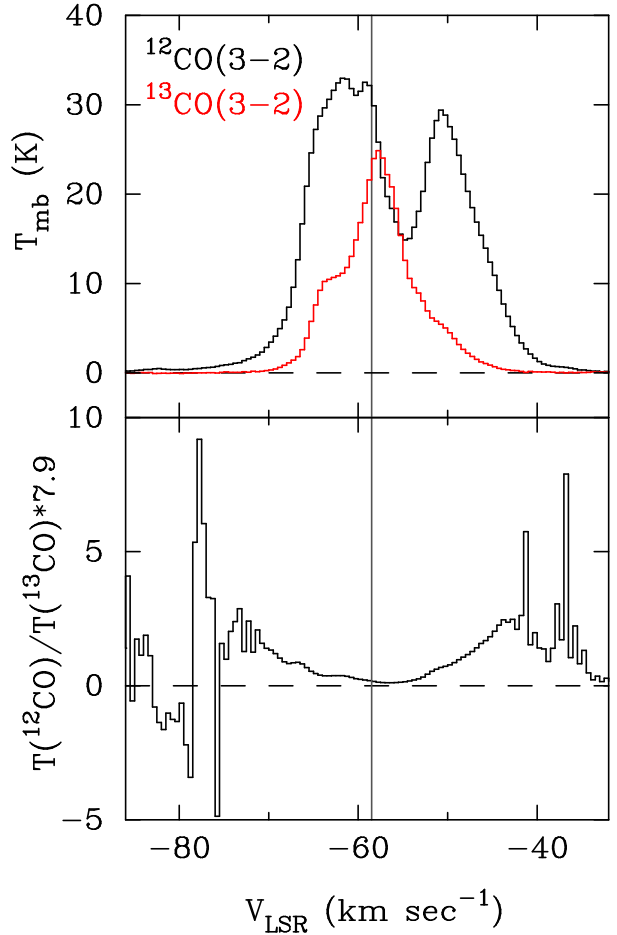}
\figcaption[]{
\label{fig-co13co}
The top panel shows long integration $^{12}$CO(3--2) and  $^{13}$CO(3--2)
spectra, the latter plotted in red, toward IRS\,1. The bottom panel shows the
ratio of T$_{mb}$($^{12}$CO) over T$_{mb}$($^{13}$CO), with $^{13}$CO multiplied
by a factor of 7.9, i.e. a ratio of 10 corresponds to optically thin
$^{12}$CO(3--2) emission. This shows that $^{12}$CO(3--2) is significantly
optically thick over the whole velocity range, where we see the high velocity
gas in $^{13}$CO, i.e. to $\sim$ -78 \kms, or $\sim$ 20 \kms\ from the systemic
velocity in the blue-shifted outflow wing. At high outflow velocities the ratio
is dominated by the noise in the spectra. The gray vertical line marks the
systemic velocity, -58.5 \kms. There are two clouds in the direction of IRS1,
one at  -58 \kms\ and the other one, which is more diffuse at $\sim$ -50 \kms.
The -58 \kms\ cloud is self absorbed in $^{12}$CO. Hence  the two clouds look
about equal in  $^{12}$CO, while the -50 \kms\ cloud is fainter in 
$^{13}$CO(3--2).
}
\end{figure}

We use the CO(3--2) map to to estimate the physical properties of the outflow, 
because it is the only outflow tracer we have, which covers the whole outflow.
We restrict the analysis to the
blue-shifted outflow lobe, because it is clearly defined with very little
contamination for the surrounding cloud. It is harder to extract information
from the counter-flow ("red"-shifted), because of the strong cloud emission and
contamination from other outflows. It is not possible to directly estimate the
inclination of the blue-shifted outflow lobe, but it appears to be  close to the
plane of the sky. Here we assume an inclination of 70\degr\ $\pm$ 10\degr. If we
use the total extent of the blue-shifted outflow, 3.6 pc, a velocity of 10 \kms{} (
see Fig~\ref{fig-co32cut10}), we get t$_{dyn}$ $\sim$ 1.3 $\times$ 10$^5$ yr
after correcting for inclination. This estimate has an uncertainty of about a
factor of two due to the uncertainty in the inclination. Usually dynamical
timescales are under-estimates, see e.g. \citet{Parker91}. When an outflow first
starts, it has to drill through dense gas, so it starts slowly. Then once the
density gets lower (and it does seek the path of least resistance), the
outflow speeds up. How long this phase takes is hard to estimate. After that it
may move with relatively constant velocity as long as the gas density is
relatively uniform, which is definitely not the case here. Nevertheless, here we
will use the dynamical time scale derived above, i.e.  1.3 $\times$ 10$^5$ yr.
For estimating physical parameters of the outflow, like mass, momentum and
kinetic energy, we use the large JCMT CO(3--2) map, which is the only data set
we have that covers the whole outflow.  Here we only analyze the northern
blue-shifted outflow lobe, because it is well defined and only moderately
affected by the surrounding molecular cloud, where as the same is not true for
the southern outflow, which is strongly affected by the dense molecular cloud
south of IRS\,1.  There may be some contribution from outflows  from embedded
lower mass protostars \citep{Qiu11}, but the IRS\,1 outflow will dominate
overall.  CO(3--2), however, is extremely optically thick, and we do not have
$^{13}$CO(3--2) maps covering more than a small portion of the outflow.  To get
an idea of the $^{12}$CO  optical depth we therefore use the deep HARP CO(3--2)
and $^{13}$CO(3--2) maps, which are centered on IRS\,1 and  cover $\sim$
100\arcsec\ $\times$ 100\arcsec\ and 80\arcsec\  $\times$ 80\arcsec\ for
$^{12}$CO(3--2) and  $^{13}$CO(3--2) respectively. These maps show that
$^{12}$CO is significantly optically thick over  all outflow velocities seen in
$^{13}$CO, or up to blue-shifted velocities of  $>$ 20 \kms\ from the systemic
velocity, here assumed to be -58.5 \kms. This is  shown in Fig.\ref{fig-co13co},
where we show the $^{12}$CO(3--2) and $^{13}$CO(3--2) spectra in the top panel
and the ratio of $^{12}$CO(3--2) to $^{13}$CO(3--2) in the bottom panel. This
figure shows that $^{12}$CO is optically thick over the velocity range were we
detect  $^{13}$CO emission, Here we have assumed a  $^{12}$C to  $^{13}$C ratio
of 79 \citep{Wilson94} and scaled the $^{13}$CO(3--2) by 7.9, i.e. the ratio
would be 10 if $^{12}$CO(3--2) is optically thin. However, $^{12}$CO(3--2) is
not only optically thick on IRS\,1, similar ratios are seen over the whole area
mapped in  $^{13}$CO(3--2), indicating that  $^{12}$CO is likely to be optically
thick over a large part of the outflow.

To estimate the outflow parameters we therefore divided up the map into three
parts. The first is a polygon encompassing the outflow around IRS\,1 to
50\arcsec\ north of the star, the second goes from there to $\sim$ 100\arcsec\
from IRS\,1 and the third one covers the tip of the outflow $\sim$ 250\arcsec\
to the north. We integrate over velocity intervals of 2.1 \kms\ for low
velocities, then gradually increasing the velocity range to 3.8 \kms. The
highest velocity range, 20.5 \kms\ from the systemic velocity is integrated over
6.4 \kms. For the area closest to IRS1 we apply the average opacity corrections
derived from the $^{12}$CO to  $^{13}$CO ratio. Since we don't know what the
optical depth is further out, we take half of the correction and at the tip of the
outflow we assume the CO is optically thin.

The IRS\,1 molecular outflow is hot. Near IRS\,1 \citet{Klaassen09,Zhu13} find
the temperature of the molecular gas is 250 K or more from analysis of
methylcyanide emission toward IRS\,1. Since one sees CO(11--10) emission up to more
than 40\arcsec\ from IRS\,1 and  strong \CII\  emission to more than 80\arcsec\
from IRS\,1, the gas must be hot even further away from IRS\,1. It is therefore
clear that strong UV radiation from IRS\,1 shining into the outflow cavity and
illuminating the gas in the outflow is capable of heating the gas over most of
the outflow. Here we have assumed a gas temperature of 100 K for our column density
estimates, which is the same temperature \citet{Klaassen11} used in their study of
outflows from high mass stars. The error resulting from the uncertainty in
temperature is relatively minor. A 50 K error in temperature only changes the
column density by 30 -- 34\%, whereas errors in the  $^{12}$CO  optical depth
over the outflow is likely to result in an uncertainty of a factor of two or
more in the mass of the outflow.  We estimate the outflow mass for optically
thin emission assuming a standard CO to H$_2$ abundance ratio of 10$^{-4}$
\citep{Mangum15}, a mean molecular weight $\mu_{\rm{H}_{2}}$ of 2.8 per hydrogen
molecule \citep{Kauffman08} and apply the opacity corrections as explained
earlier in this section.  By summing over all the velocity intervals from -61.4
\kms\ to -82.2 \kms\  we get a total mass of the blue-shifted outflow lobe of 45.5
\Msun. More than 90\% of the mass is the area extending up to 50\arcsec\ from
IRS\,1. To calculate momentum, P, momentum flux, F, and kinetic energy, E$_k$,
we use the formulae presented in \citet{Choi93}, although we correct for
inclination assuming an outflow inclination of 70\degr. For the blue-shifted
outflow we get P =  290 \Msun{}\kms,  E$_k$ = 1.9 $\times$ 10$^{40}$ J, and F =
0.4 $\times$ 10$^{-2}$ \Msun{} km s$^{-1}$ yr$^{-1}$. If we assume that the
outflow is momentum conserved, the mass in the red-shifted outflow is likely to
be similar, since the outflow velocities are about the same for both the blue-
and the red-shifted outflow. The same is true for the other outflow parameters.
If we compare our results to \citet{Qiu11} over the same velocity interval as
they used, our mass estimate, 91  \Msun\ agrees  quite well with their
estimate, $\sim$ 100 \Msun. We can also estimate the mass of the outflow in
\CII\ emission. To compute the \CII\ column density we use equation 26 in
\citet{Goldsmith12} assuming that the \CII| emission is optically thin with an
excitation temperature of 100 K. If we  adopt 1.2 $\times$ 10$^{-4}$ for the
C$^+$/H abundance ratio  \citep{Wakelam08}, see also \citet{Ossenkopf15}, we get
a total mass of the \CII\ outflow of 18 \Msun, by integrating the blue-shifted \CII\
emission from -74 -- -61.5 \kms. Here we have corrected the outflow mass for
contribution from IRS\,2, which we know has strong blue-shifted \CII\ emission,
see Section~\ref{sect-CII}, by placing a 20\arcsec\ aperture centred on IRS\,2.
The mass of gas ionized by IRS\,2 is $\sim$ 3 \Msun.

The total mass of the blue-shifted outflow is therefore $\sim$64 \Msun, and if we assume 
that the mass in the red-shifted outflow is the same, the total outflow mass could be as high as 130 \Msun, 
which is still less than the mass of the core of IRS,1, which we estimated to be $\sim$ 490 \Msun, 
Section~\ref{sect-nature}. Compared to low-mass outflows, the IRS\,1 outflow is much more massive, but the mass is similar to other
high mass star outflows \citep{Beuther02}. There are examples of high mass star outflows,
where the outflow mass even exceed the core mass \citep{Shepherd98,Qiu09}.
Earlier we estimated a dynamical time scale of 1.3
$\times$ 10$^5$ yr, which would give us a mass outflow rate of 1.0 10$^{-3}$
\Msun{}yr$^{-1}$, which agrees reasonably well with the observed mass accretion rate.

\section{Discussion}

The IRS\,1 -- 3 region, which borders the \ion{H}{2} region NGC\,7538, is the
center of a young star cluster with more than 150 young stars and protostars
\citep{Mallick14,Sharma17}. The IRS\,1 - 3 cluster is almost certainly a case of
star formation triggered by the expanding NGC\,7538 \ion{H}{2} region, although
it has not been conclusively proven to be the case
\citep{Balog04,Puga10,Sharma17}. IRS\,1 is the most massive star in the cluster
and it is still  heavily accreting with an accretion rate in the range 10$^{-3}$
to a few times 10$^{-4}$ \rm M$_{\odot}$/yr \citep{Sandell09, Klaassen11, Qiu11,
Beuther12, Zhu13}, which quenches the expansion of an  \ion{H}{2} region. It is
almost certainly surrounded by an accretion disk, although there is no firm
evidence of such a disk, because the emission from the {\bf ionized} jet powered by IRS\,1 is so strong in the millimeter and submillimeter
regime, that it completely hides the much fainter accretion disk. 

In this paper we have
analyzed high angular resolution ($\sim$ 0\ptsec1)  CARMA observations at 1.3 mm
which conclusively shows that none of the previously proposed companions to IRS\,1 are high mass stars. It could still be a binary, but then the secondary must be within 30 AU,
if the binary components have roughly equal brightness or if the secondary has a
later spectral type. In a way this is somewhat surprising since most massive
stars are close binaries or multiple systems \citep{Sana12}. A single star is, however,
consistent with the observed luminosity of the star, $\geq$ 1.0 10$^5$ \Lsun,
for which the lower limit corresponds to the luminosity of a single O7 star.  If the spectral type is as early as O5.5, 
the secondary would have to be a later spectral type. For a spectral type of O7 – O5.5, 
the stellar mass would be in the range 30 -- 50  \Msun{} \citep{Davies11}.

What makes IRS\,1 so extraordinary is that it is an extremely young O-star,
which has such a high accretion rate that it quenches the formation of an
\ion{H}{2} region. The FUV-radiation by the star can only escape in the polar
regions, where it excites a "jet-like" ionized outflow with velocities in excess of
250 \kms \citep{Gaume95}. In this paper we have also shown that the strong FUV
radiation from IRS\,1 heats molecular gas inside the outflow lobes and ionizes
carbon, which is why we see high velocity \CII\ 158 $\mu$m emission up to $\sim$
100\arcsec\  (1.3 pc) north of IRS\,1 in the blue-shifted outflow lobe. Most of
this gas is still molecular and seen in CO(3--2), with similar morphology and
velocity structure, see Figs. \ref{fig-co32cut10} and  \ref{fig-ciicut10}.

We can use the jet model by \citet{Reynolds86} to estimate the turnover frequency and the mass loss rate of the ionized jet, i.e., his Eqs. 18 and 19.
In his paper \citet{Reynolds86} used IRS\,1 as an example to illustrate of how his model could be applied. Since we now have more accurate data and knowledge of IRS\,1, we can expect more robust results. From the VLA data in \citet{Sandell09} we estimate a collimation angle of $\sim$ 15\degr.
We further assume a wind velocity of 500 \kms,  which is about what one would expect
for a wind from an O-star \citep[see e.g.,][]{Johnston13,Sanna16,Rosero19}, and that the jet is fully ionized. We determined the spectral index $\alpha$ = 0.87 and estimated the inclination of the jet to be $\sim$ 70\degr. Since we know the spectral index, we can estimate the rest of the jet parameters from Eqs. 15 and 17 in \citet{Reynolds86}, see also \citet{Sanna18}, who analyzed radio jets from 33 radio jets.  If we assume a launching radius of 20 AU, i.e. the radius at which we estimated the jet to become optically thin (Section~\ref{sect-nature}),  we get a turn over frequency of 620 GHz, which agrees very well with our estimate from the fall off in size, 700 - 1500 GHz.
If the jet is launched closer to the star, the turnover frequency would be higher. Reynolds’ equation 19 now gives a mass loss rate of 1.2 10$^{-4}$
\Msun{}yr$^{-1}$. Although this mass loss rate is 3 - 10 times higher than we found
for any jet in the literature \citep[see e.g.,][]{Johnston13,Zhang19}, it appears quite plausible and suggests the jet now has enough momentum to drive the IRS\,1 molecular outflow.

The core accretion models by \citep{Tanaka16} predict that a high-mass star starts with
a small thermal jet confined by the outflow from the protostar. The photo ionized region is confined by the magneto-hydrodynamically  driven outflow. The models predict that almost the
whole of the outflow is fully ionized in 10$^3$ - 10$^4$ yr after the initial \ion{H}{2}
region is formed. The models do not describe the phase we see for IRS\,1, i.e.,
when the star’s ionizing flux is turned on and the accretion rate is still so high, that the ionizing flux can only escape in the polar regions, i.e., into the outflow, creating the ionized jet we now see. \citet{Tanaka17} do address the onset of ionization, but their models only go up to a stellar mass of 16 \Msun,
and are therefore not applicable to IRS\,1, which is much more massive.

We have confirmed that IRS\,1 drives a large
parsec scale outflow, which in the north extends to a projected distance of 3.6
pc from IRS\,1, giving a dynamical timescale of the outflow of 1.3 $\times$
10$^5$ years, see Section~\ref{sect-Outflow_prop}, which is probably an
underestimate. In their review of massive star formation, \citet{Zinnecker07}
argued that high mass star formation is not a scaled up version of low mass star
formation. It requires a cluster environment with a much larger mass reservoir
and higher pressure than in a low mass star forming environment. However, in the
very earliest stages of the formation of a high mass star it may be very
difficult or impossible to judge whether the protostar will form a low mass star
or evolve into a high mass star. Based on the dynamical timescale, IRS\,1 most
likely started as a much lower mass star. Although there is no way to estimate
the accretion rate in the past, it probably started with a much lower accretion
rate. The accretion rate increased in time, when the mass of the core in
which it was embedded grew in size, i.e., essentially what is often called competitive 
accretion, but perhaps better called runaway accretion \citep{Zinnecker82} or hierarchical accretion \citep{Larson78}. 
IRS\,1 will continue to grow in mass until the accretion rate drops, so that it can
no longer quench the \ion{H}{2} region. When this happens the \ion{H}{2} will
rapidly expand and disperse the material around it. There is even two other
examples of  high-mass stars in NGC\,7538 that are going through the same process,
IRS\,9 \citep{Sandell05,Rosero19} and NGC\,7538 South \citep{Sandell10}. Both IRS\,9 
and NGC\,7538 South are younger than IRS\,1 and have similar
accretion rates as IRS\,1. Both have currently weak radio jets and luminosities suggesting that they are still early B-stars.
Given the current accretion rates, they will rapidly evolve into O-stars like
IRS\,1. 

\section{Summary}

NGC\,7538~IRS\,1 is a very young O-star at a distance of 2.65 kpc, which is
still so heavily accreting, that the accretion prevents the formation of an
\ion{H}{2} region. The new observational data presented in this paper provide
new insight on IRS\,1 and suggest that at least some O-stars form the same way
as low mass stars, i.e. with an accretion disk and driving an outflow, but with
much higher accretion rates.

The data presented in this paper confirm the model proposed by \citet{Sandell09},
who showed that the free-free emission from IRS\,1 is dominated by a collimated
ionized wind driving an ionized north-south jet. Re-analysis with new data from
the literature and from this paper we find that the radio SED follows a power
law with a spectral index, $\alpha$ = 0.87 $\pm$ 0.03, which is very similar to
what \citet{Sandell09} found. Only at frequencies higher than 300 GHz one can
start seeing some excess emission due to thermal dust emission. The size of the
free-free core falls off as $\nu^{-0.92 \pm 0.02}$, which is consistent with a
jet \citep{Reynolds86}.  If we assume that the jet is launched from a radius of
10 -- 20 AU, this fit predicts that the free-free emission will not become
completely optically thin until somewhere between 700 -- 1500 GHz. Mapping of
the outflow from IRS\,1 in \CII\ and CO(11--10) confirm that strong FUV
radiation leaks out in the polar regions of the IRS\,1 core (or disk) ionizing
carbon and heating up the molecular gas.

We find no evidence for IRS\,1 being a binary or triple system as has been
suggested in the past. Analysis of high angular resolution ($\sim$ 0\ptsec1)
CARMA observations at 1.3 mm show a barely resolved single source. It could
still be a binary, but if it is, the separation between the two stars must be
less than 30 AU if the binary components have roughly equal brightness. With our
spatial resolution and sensitivity we would have easily detected the two binary
components proposed by \citet{Beuther17}.

We find that IRS\,1 drives a massive parsec scale outflow, which in the north
has sculpted a cavity extending to 3.6 pc from IRS\,1. To the south it is more
difficult to trace the outflow, because it penetrates the massive molecular
cloud south of NGC\,7538. IRAC images suggest that it terminates at $\sim$ 2.6
pc from IRS\,1. At high velocities the northern outflow is well aligned with the
free-free outflow. We trace high velocity cloudlets inside the outflow cavity to
at least 80\arcsec\ north of IRS\,1 in both CO(3--2) and \CII.  Since CO and
\CII\ show the same morphology and velocity structure they originate in the same
gas, i.e. \CII\ traces the outer layers of the gas ionized by the strong FUV
field from IRS\,1 shining into the cavity. We do not see these cloudlets in
CO(11--10), which is a tracer of high density hot gas, but we see CO(11--10) all
the way down to NGC\,7538~South. This gas can only be heated by FUV radiation
from IRS\,1. We estimate a dynamical time scale of the outflow of 1.3 $\times$
10$^5$ years and a total outflow mass of 130 \Msun, which would give us a mass
outflow rate of 1.0 $\times$ 10$^{-3}$ \Msun{}yr$^{-1}$. This agrees quite well with the
observed accretion rate, which is in the range  10$^{-3}$ to a few times
10$^{-4}$ \Msun{}yr$^{-1}$, especially since the dynamical time scale is likely
to be underestimated.

We find that the O-star IRS\,1 was formed in a similar way as low-mass stars,
i.e., with an accretion disk and driving an outflow, but in a much denser
environment and with a much higher accretion rate. It is likely that many other
high mass stars may form the same way, although other formation scenarios are
also possible. In the NGC\,7538 molecular cloud there are also two other high-mass
star: IRS\,9 and NGC\,7538 South, which both appear to be younger analogues of IRS\,1.
Therefore at least some high-mass stars form in a similar fashion as low mass stars.

\acknowledgements 
Based in part on observations made with the NASA/DLR Stratospheric Observatory
for Infrared Astronomy (SOFIA). SOFIA is jointly operated by the Universities
Space Research Association, Inc. (USRA), under NASA contract NNA17BF53C, and the
Deutsches SOFIA Institut (DSI) under DLR contract 50 OK 0901 to the University
of Stuttgart. The development
of GREAT (German REceiver for Astronomy at Terahertz frequencies) was financed
by the participating institutes, by the Federal Ministry of Economics and
Technology via the German Space Agency (DLR) under Grants 50 OK 1102, 50 OK 1103
and 50 OK 1104 and within the Collaborative Research Centre 956, sub-projects D2
and D3, funded by the Deutsche Forschungsgemeinschaft (DFG). Support for CARMA 
construction was derived from the Gordon and
Betty Moore Foundation, the Kenneth T. and Eileen L. Norris Foundation, the
James S. McDonnell Foundation, the Associates of the California Institute of
Technology, the University of Chicago, the states of California, Illinois, and
Maryland, and the National Science Foundation. Ongoing CARMA development and
operations were supported by the National Science Foundation under a cooperative
agreement, and by the CARMA partner universities. We acknowledge support from
Onsala Space Observatory for  the  provisioning of its facilities/observational
support. The Onsala Space Observatory national research infrastructure is funded
through Swedish Research Council grant No 2017-00648. We thank Dr. Hans Zinnecker for
constructive criticism of the paper. We thank the
referee, for comments  which improved this manuscript.

\newpage
\appendix
\renewcommand\thefigure{\thesection.\arabic{figure}}   
\section{Appendix A: Channel maps}
\setcounter{figure}{0} 
Below we present channel maps of the JCMT CO(3--2) map and the GREAT \CII, and CO(11--10) maps discussed in the main body of the paper.

\begin{figure*}[b]
\includegraphics[width=\textwidth, angle=-90]{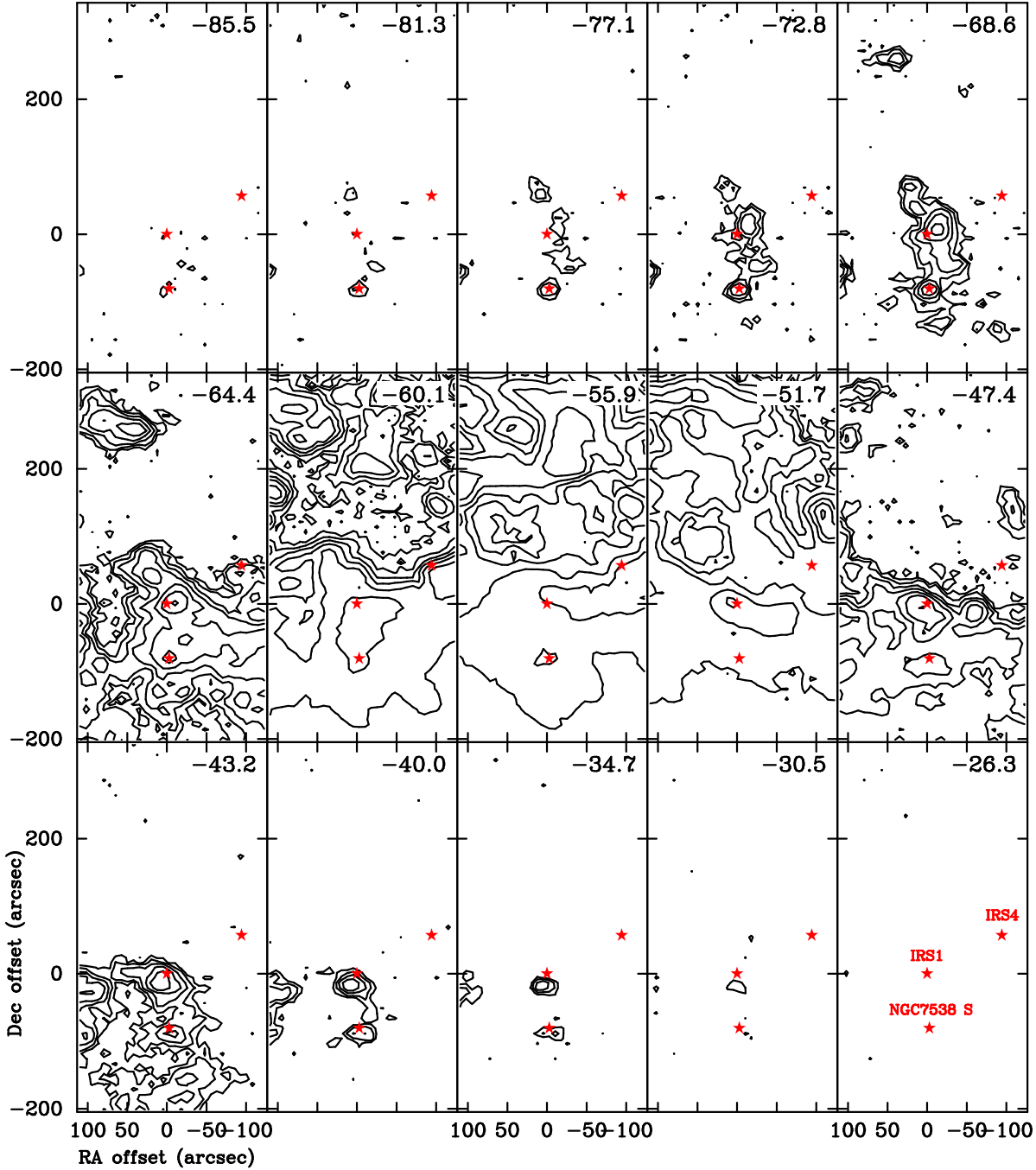}
\figcaption[]{
\label{fig-CO32_channels}
Channel maps of CO(3--2) averaged over 4.2 \kms\ wide velocity intervals. The
center velocity of each panel is indicated in the top right corner of each
panel. The emission is plotted with eight logarithmic contours from 0.35 K to 26
K. The red stars in each panel are IRS\,1, NGC\,7538 South and IRS\,4. These are
labelled in the bottom right panel (velocity -26.3 \kms{}). At high blue- and
red-shifted velocities one only sees emission from IRS\,1, and NGC\,7538 South.
IRS\,4 is not associated with any outflow. At velocities around -60 \kms\ we can
see the boundary between the NGC\,7538 \ion{H}{2} region and the molecular could
south of it. Emission in the channel at -47 \kms\ also appears to trace the
boundary of the \ion{H}{2} region.
}
\end{figure*}

\begin{figure*}
\includegraphics[width=\textwidth, angle=-90]{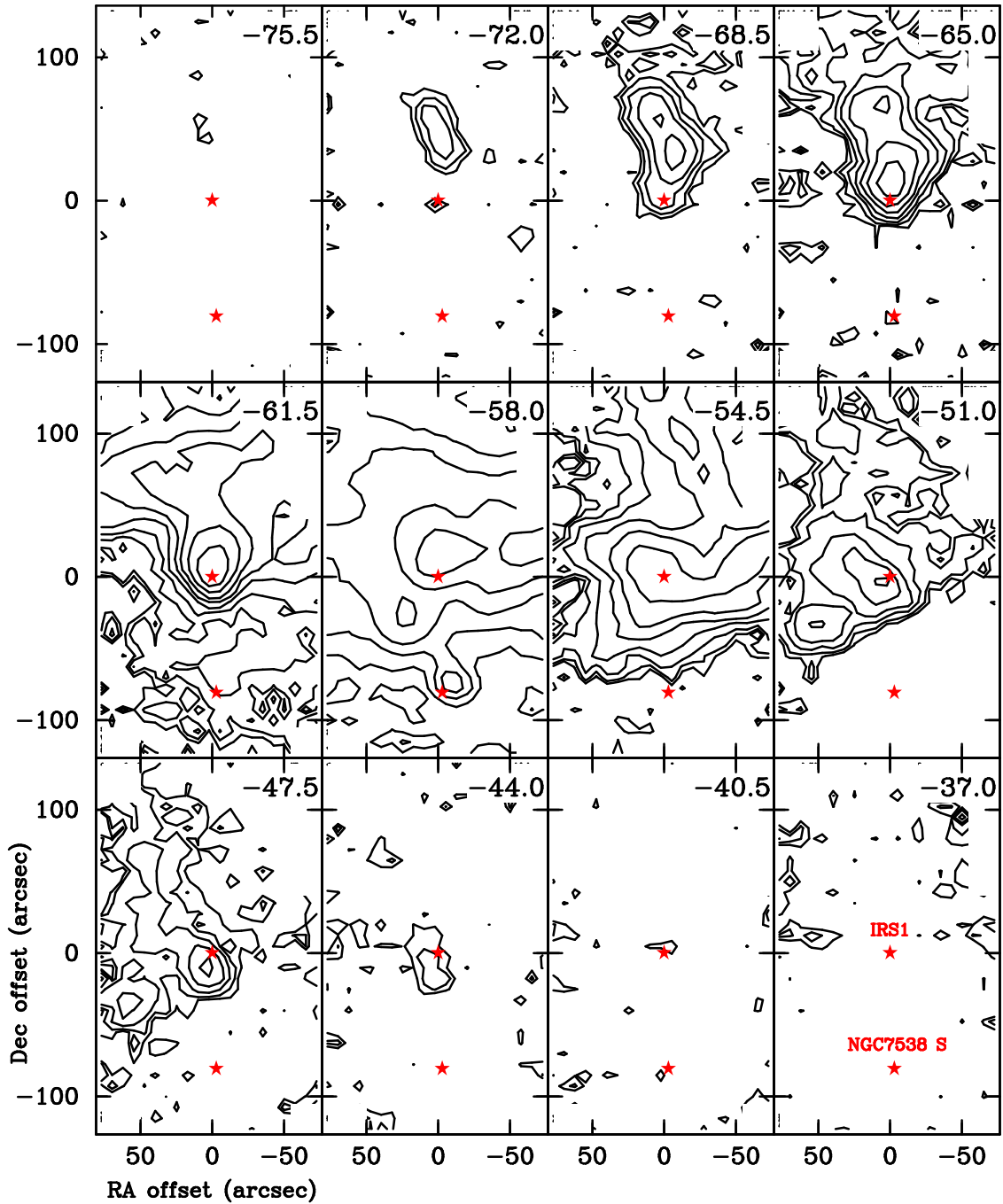}
\figcaption[]{
\label{fig-CII_channels}
Channel maps of \CII\ averaged over 3.5 \kms\ wide velocity intervals. The
center velocity of each velocity interval  is indicated in the top right corner
of each panel. The emission is plotted with ten logarithmic contours from 0.4 K
to 40 K. The red stars in each panel are IRS\,1, NGC\,7538 South. These are
labelled in the bottom right panel (velocity -37.0 \kms{}). One can see
widespread PDR emission from both the -58 \kms\ and -50 \kms\ clouds. NGC\,7538
South is not associated with any \CII\ emission. The compact emission NW of
NGC\,7538 South appears centered on IRS\,11, another young early B-star in the
cloud.
}
\end{figure*}

\begin{figure}
\includegraphics[width=8.5cm, angle=-90]{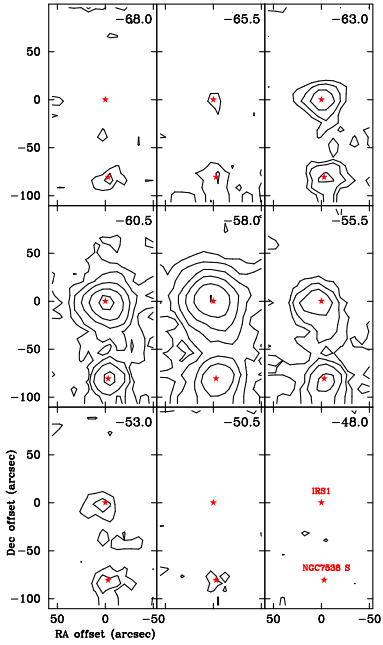}
\figcaption[]{
\label{fig-CO11_channels}
Channel maps of CO(11-10) averaged over 2.5 \kms\ wide velocity intervals. The
emission is plotted with six logarithmic contours from 0.5 K to 24 K. The center
velocity of each velocity interval is indicated in the top right corner of each
panel. The red stars in each panel are IRS\,1, NGC\,7538 South. These are
labelled in the bottom right panel. There is strong CO(11--10) high velocity
emission from both IRS\,1 and NGC\,7538 South. At low velocities, panels -60.5
\kms\ to -55.5 \kms, the CO(11--10) emission extends all the way from IRS\,1 to
NGC\,7538 South. The emission extending to the west from IRS\,1, which is seen
in panels -60.5 and -58 \kms\ is almost certainly associated with the NGC\,7538
PDR rim.
}
\end{figure}


\begin{thebibliography}{}

\bibitem[Akabane \& Kuno(2005)]{Akabane05}
   Akabane, K., \& Kuno, N., 2005, \aap, 431, 183

\bibitem[Akabane et al.(1992)]{Akabane92}
   Akabane, K., Tsunekawa, S., Inoye, M., et al.  1992, \pasj, 44, 435

\bibitem[Alonso-Mart\'ines et al.(2017)]{Alonso17}   
   Alonso-Mart\'ines, M., Riviere-Marichalar, P. Meeus, G., et al.  2017, 
   \aap, 603, A138
   
\bibitem[Balog et al.(2004)]{Balog04}
     Balog, Z., Kenyon, S. L., Lada, E. A., et al. 2004, \aj, 128, 2942
     
\bibitem[Beuther et al.(2017)]{Beuther18}
     Beuther, H., Mottram, J. C., Ahmadi, A., et al.  2018, \aap, 617, A100

\bibitem[Beuther et al.(2017)]{Beuther17}
     Beuther, H., Linz, H., \& Henning, Th., et al.   2017, \aap,  605, A61 

\bibitem[Beuther, Linz \& Henning(2012)]{Beuther12}
     Beuther, H., Linz, H., \& Henning, Th.  2012, \aap, 543, A88
     
\bibitem[Beuther et al.(2002)]{Beuther02}
     Beuther, H., Schilke, P., Sridharan, T. K., et al.  2002, 
     \aap, 383, 892     
     
\bibitem[Buckle et al.(2009)]{Buckle09} 
        Buckle, J. V., Hills, R. E., Smith, H., et al.  2009, 
         \mnras, 399, 1026 
         
\bibitem[Campbell(1984)]{Campbell84}
        Campbell, B.  1984, \apjl, 282, L27
   
\bibitem[Choi, Evans II \& Jaffe(1993)]{Choi93}
        Choi, M., Evans II, N. J., \& Jaffe, D. T.  1993, \apj,  417, 624 
   
\bibitem[Corder(2008)]{Corder08}
        Corder, S.  2008, PhD thesis, Caltech
   
\bibitem[Crampton, Georgelin \& Georgelin(1978)]{Crampton78}
       Crampton, D., Georgelin, Y. M., \& Georgelin, Y. P.  1978, \aap, 66,1
 
\bibitem[Davies et al.(2011)]{Davies11}
      Davies, B., Hoare, M. G., Lumsden, S. L., et al.  2011, \mnras, 416, 972

   
\bibitem[Davis et al.(1998)]{Davis98}
      Davis, C. J., Moriarty-Schieven, G., Eisl\"offel, J., et al.  1998,
      \aj, 115,1118

\bibitem[De Buizer \& Minier(2005)]{DeBuizer05}
      De Buizer, J. M., \& Minier, V. 2005, \apj, 628, L151

\bibitem[De Buizer(2003)]{DeBuizer03}
      De Buizer, J. M. 2003, \mnras, 341, 277

\bibitem[De Pree et al.(1994)]{DePree94}
     De Pree, C. G., Goss, W. M.,  Palmer, P., et al.  1994,
     \apj, 428, 670
     
\bibitem[Fallscheer et al.(2013)]{Fallscheer13}
      Fallscheer, C., Reid, M. A., Di Francesco, J., et al.  2013, \apj,
      773, 102
           
\bibitem[Fischer et al.(1985)]{Fischer85}
     Fischer, J., Sanders, D. B.,  Simon, M., et al.  1985, \apj, 293, 508
     
\bibitem[Flower \& Pineau des For\^ets(2010)]{Flower10}
     Flower, D. R., \& Pineau des For\^ets, G. 2010,\mnras, 406, 1745
      
\bibitem[Franco-Hern\'{a}ndez \& Rodr\'{\i}guez(2004)]{Franco-Hernandez04}
    Franco-Hern\'{a}ndez, R., \& Rodr\'{\i}guez, L. F.  2004, \apj, 604, L105
    
\bibitem[Frau et al.(2014)]{Frau14}
    Frau, P., Girart, J. M., Zhang, Q., et al. 2014, \aap, 567, A116
    
\bibitem[G\'alvan-Madrid et al.(2010)]{Galvan10}
     Galv\'an-Madrid, R., Montes, G., Ram\'{\i}rez, E. A., et al.  2010,
     \apj, 713, 423
    
\bibitem[Gaume et al.(1995)]{Gaume95}
   Gaume, R. A., Goss, W. M., Dickel, H. R., et al. 1995, \apj, 
   438, 776
   
\bibitem[Goddi, Zhang \& Moscadelli(2015)]{Goddi15}   
   Goddi, C., Zhang, Q., \& Moscadelli, L.  2015, \aap, 573, A108
   
\bibitem[Goldsmith et al.(2012)]{Goldsmith12}
    Goldsmith, P. F., Langer, W. D., Pineda, J. L., et al.  2012, 
    \apjs, 203,13
   
\bibitem[Green et al.(2013)]{Green13}
    Green, J. D., Evans II, N. J., J\o rgensen, J. K., et al.  2013,
    \apj, 770, 123
    
\bibitem[Hackwell, Grasdalen \& Gehrz(1982)]{Hackwell82}
   Hackwell, J. A., Grasdalen, G. L., and Gehrz, R. D.  1982, \apj,
   252, 250
     
\bibitem[Hardebeck(1971)]{Hardebeck71}
     Hardebeck, E.  1971, \apj, 170, 281
     
\bibitem[Herter et al.(2013)]{Herter13}     
      Herter, T. L., Vacca. W. D., Adams, J. D., et al.  2013 
      \pasp, 125, 1393
      
\bibitem[Herter et al.(2012)]{Herter12}
     Herter, T. L., Adams, J. D., De Buizer, J. M., et al. 2012, 
     \apjl, 749, L18         
\bibitem[Heyminck et al.(2012)]{Heyminck12} 
     Heyminck, S., Graf, U. U., G\"usten, R., et al.   2012, 
     \aap, 542, L1           

\bibitem[Hollenbach \& Tielens(1999)]{Hollenbach99}
      Hollenbach, D. J., \& Tielens, A. G. G. M.  1999, 
      Rev of Modern Physics, 71, 173
      
\bibitem[Johnston et al.(2013)]{Johnston13}
      Johnston, K. J., Shepherd, D. S., Robitaille, T. P., et al.  2013, 
      \aap, 551, A43

\bibitem[Kameya  et al.(1989)]{Kameya89}
      Kameya, O., Hasegawa, T. I., Hirano,, N., et al.  1989, 
      \apj, 339, 222
      
\bibitem[Kauffman et al.(2008)]{Kauffman08}
      Kauffman, J., Bertoldi, F., Bourke, T. L., et al. 2008, \aap, 487, 993

\bibitem[van Kempen et al.(2010)]{Kempen10}
      van Kempen, T.A., Kristensen, L. E., Herczeg, G. J., et al.  2010, 
      \aap, 518, L121

\bibitem[Keto et al.(2008)]{Keto08}
     Keto, E., Zhang, Q., \& Kurtz, S.  2008, \apj, 672, 423
   
\bibitem[Klaassen et al.(2011)]{Klaassen11}
     Klaassen, P. D., Wilson, C. D., Keto, E. R., et al.  2011, \aap, 530, A53
 
 \bibitem[Klaassen et al.(2009)]{Klaassen09}
     Klaassen, P. D., Wilson, C. D., Keto, E. R., et al.  2009, 
     \apj, 703, 1308

\bibitem[Klein et al.(2012)]{Klein12}
         Klein,  B., Hochg\"urtel, S., Kr\"amer, I., et al.  2012, 
         \aap,  542, L3               
\bibitem[Kraus et al.(2006)]{Kraus06}
      Kraus, S., Balega, Y., Elitzur, M., et al.  2006, \aap, 455, 521
   
\bibitem[Kurtz  \& Franco(2002)]{Kurtz02}
     Kurtz, S., \& Franco, J. 2002,  RevMexAA, 12,  16

\bibitem[Larson(1978)]{Larson78}
     Larson, R. B.   1978, \mnras, 184, 69

\bibitem[Lugo,  Lizano \& Garay(2004)]{Lugo04}
     Lugo, J., Lizano, S., \&  Garay, G.  2004, \apj, 614, 807
     
\bibitem[Mallick et al.(2014)]{Mallick14} 
      Mallick, K. K., Ojha, D. K., Tamura, M., et al.  2014, 
      \mnras, 443, 3218
      
\bibitem[Mangum \& Shirley(2016)]{Mangum15}
       Mangum, J. G., \& Shirley, Y. L.  2015, \pasp, 127, 266
       
\bibitem[Martin(1973)]{Martin73} 
      Martin, A. H. M.   1973, \mnras,  163,  141  
     
\bibitem[Minier, Booth \& Conway(1998)]{Minier98}
      Minier, V., Booth, R. S., \& Conway, J. E. 1998, \aap, 336, L5
      
\bibitem[Minier, Booth \& Conway(2000)]{Minier00}
      Minier, V., Booth, R. S., \& Conway, J. E. 2000, \aap, 362, 1093
      
\bibitem[Moreno \& Chavarr\'{i}a-K(1986)]{Moreno86}
     Moreno, M. A., and Chavarr\'ia-K, C.  1986, \aap, 161, 130 
       
\bibitem[Moscadelli et al.(2008)]{Moscadelli08}
     Moscadelli,  L., Reid, M. J., Menten, K. M., et al.  2008,  
     \apj, 693, 406

\bibitem[Moscadelli \& Goddi(2014)]{Moscadelli14}
     Moscadelli,  L., \& Goddi, C.  2014, \aap, 566, A150
     
\bibitem[Mueller et al.(2002)]{Mueller02}
      Mueller, K. E., Shirley, Y. M., Evans II, N. J., et al.   2002,
      \apjs, 143, 469
      
\bibitem[Ossenkopf et al.(2015)]{Ossenkopf15}
     Ossenkopf, V., Koumpia, E., Okada, Y., et al. 2015, \aap, 580, A83 
      
\bibitem[Parker, Padman \& Scott(1991)]{Parker91}      
     Parker, N. D., Padman, R., \& Scott, P. F.  1991, \mnras, 252, 442. 
     
\bibitem[Pestalozzi et al.(2004)]{Pestalozzi04}
     Pestalozzi,  M. R., Elitzur, M., Conway, J. E., \& Booth, R. S.
     2004, \apj, 603, L113
     
\bibitem[Pestalozzi, Elitzur \& Conway(2009)]{Pestalozzi09}
     Pestalozzi,  M. R., Elitzur, M., \& Conway, J. E.  2009, 
     \aap, 501, 999
     
\bibitem[Podio et al.(2012)]{Podio12}
      Podio, L., Kamp, I., Flower, D., et al. 2012, \aap, 545, A44
      
\bibitem[Puga et al. (2010)]{Puga10}  
      Puga, E., Mar\'in-Franch, A., Najarro, F., et al.  2010, 
      \aap, 517, A2
     
\bibitem[Qiu, Zhang \& Menten(2011)]{Qiu11}    
      Qiu, K., Zhang, Q., \& Menten, K. M.  2011, \apj, 728, 6
      
\bibitem[Qiu et al.(2009)]{Qiu09}    
      Qiu, K., Zhang, Q., Wu, J., et al.  2009, \apj, 696, 66      
      
\bibitem[Read(1980)]{Read80}
      Read, P. L. 1980, \mnras, 192, 11

\bibitem[Reynolds(1986)]{Reynolds86}
     Reynolds, S. P.   1986,  \apj, 304, 713
     
\bibitem[Risacher et al.(2016)]{Risacher16} 
     Risacher, C., G\"usten, R., \& Stutzki, J., et al.  2016, 
        \aap, 595, A34

\bibitem[Rodr\'{\i}guez, Anglada \& Curiel(1999)]{Rodriguez99}
      Rodr\'{\i}guez, L. F., Anglada, L., \& Curiel, S.  1999, 
      \apjs, 125, 427

\bibitem[Rosero et al.(2019)]{Rosero19} 
      Rosero, v., Tanaka, K. E. I., Tan, J. C., et al.  2019, \apj, 873,20

 \bibitem[Sana et al.(2012)]{Sana12}
     Sana, H., de Minck, S. E., de Koter, A., et al. 2012, 
     Science, 337, 44
     
 \bibitem[Sanna et al.(2018)]{Sanna18}
      Sanna, A. Moscadelli, L., Goddi, C., et al.  2018,
      \aap, 619, A107
      
\bibitem[Sanna et al.(2016)]{Sanna16}
      Sanna, A. Moscadelli, L., Cesaroni, R., et al.  2016,
      \aap, 596, L2
   
\bibitem[Sandell et al.(2012)]{Sandell12}
      Sandell, G., Wright, M., Zhu, L., et al.   2012, ALMA/NAASC 2012
      Workshop: Outflows, Winds and Jets
      (https://science.nrao.edu/facilities/alma/naasc-workshops/jets2012)
      
\bibitem[Sandell \& Wright(2010)]{Sandell10}
      Sandell, G., \&Wright, M.  2010, \apj, 715, 919           
      
\bibitem[Sandell et al.(2009)]{Sandell09}
      Sandell, G., Goss, W. M., Wright, M., \& Corder, S.  2009, 
      \apj, 699, L31        
         
\bibitem[Sandell, Goss \& Wright(2005)]{Sandell05}
    Sandell, G., Goss, W. M.,  \& Wright, M.  2005, \apj, 621, 839          
      
\bibitem[Sandell \& Sievers(2004)]{Sandell04}
       Sandell, G., \& Sievers, A.  2004, \apj, 600, 269 
       
\bibitem[Sault et al.(1995)]{Sault95}
         {Sault}, R.~J., {Teuben}, P.~J., \& {Wright}, M.~C.~H. 1995,
         in Astronomical Society of the Pacific Conference Series, Vol.~77,
         Astronomical Data Analysis Software and Systems IV, ed. R.~A. {Shaw},
         H.~E. {Payne}, \& J.~J.~E. {Hayes}, 433--+       
       
\bibitem[Sewilo et al.(2004)]{Sewilo04}
        Sewilo, M., Churchwell, E., Kurtz, S., et al. 2004, \apj, 605,285        
           
\bibitem[Scoville et al.(1986)]{Scoville86}
   Scoville, N. Z., Sargent, A. I., Claussen, M. J., et al.  1986, 
   \apj, 303, 416

\bibitem[Sharma et al.(2017)]{Sharma17}
     Sharma, S., Pandey, A. K., Ojha, D. K., et al. 2017, \mnras, 467, 2965
     
\bibitem[Shepherd et al.(1998)]{Shepherd98}     
     Shepherd, D., Watson, A. M., Sargent, A. I., et al. 1998, \apj, 507, 861
     
\bibitem[Surcis et al.(2011)]{Surcis11}  
    Surcis, G., Vlemmings, W. H. T., Torres, R. M., et al.  2011, \aap, 533, A47

 \bibitem[Tanaka(2017)]{Tanaka17}
      Tanaka, K. E. I., Tan, J. C., Staff, J. E., et al.  2017, \apj, 849, 133
      
\bibitem[Tanaka,Tan \& Zhang(2016)]{Tanaka16}
     Tanaka, K. E. I., Tan, J. C., \& Zhang, Y.  2016, \apj, 818, 52
 
     
\bibitem[Tielens \& Hollenbach(1985)]{Tielens85}
     Tielens, A. G. G. M., \& Hollenbach, D. J.  1985, \apj, 291, 722
     
\bibitem[Thronson \& Harper(1979)]{Thronson79}
    Thronson, H. A., Jr, \& Harper, D. A.  1979, \apj, 230, 133
    
\bibitem[Visser et al.(2012)]{Visser12}   
    Visser, R., Kristensen, L. E., Bruderer, S., et al   2012, \aap, 537, A55

\bibitem[Wakelam \& Herbst(2008)]{Wakelam08} 
    Wakelam, V., \& Herbst, E. 2008, \apj, 680, 371
    
\bibitem[Wiesemeyer, Thum \& Walmsley(2004)]{Wiesemeyer04}
    Wiesemeyer, H., Thum, C., \& Walmsley, C. M.  2004, \aap, 428, 479
    
\bibitem[Willner(1976)]{Willner76}
    Willner, S. P. 1976, \apj, 206, 728
    
\bibitem[Wilson \& Rood(1994)]{Wilson94}    
        Wilson, T. L., \& Rood, R. 1994, ARA\&A, 32, 191  
      
 \bibitem[Wright et al.(2014)]{Wright14}
       Wright, M. C. H., Hull, C.L. H., Pillai, T., et al.  2014, \apj, 796, 112
     
\bibitem[Wynn-Williams, Becklin, \& Neugebauer(1974)]{Wynn-Williams74}
   Wynn-Williams, C. G., Becklin, E. E., \& Neugebauer, G.  1974,
   \apj, 187, 473 

\bibitem[Yang et al.(2010)]{Yang10}
     Yang, B., Stancil, P. C., Balakrishnan, N., et al. 2010 \apj, 218, 1062
     
\bibitem[Young et al.(2012)]{Young12} 
         Young, E.~T., Becklin, E.~E., Marcum, P.~M., et al.\ 2012, 
         \apjl, 749, L17 
         
\bibitem[Zhang et al.(2019)]{Zhang19}
       Zhang, Y., Tanaka, K. E. I., Tan, J. C., et al. 2019, \apjl, 886, L4        
   
\bibitem[Zhu et al.(2013)]{Zhu13}
       Zhu, L., Zhao, J.-H., Wright, M. C. H., et  al.  2013, \apj, 779, 51
       
\bibitem[Zhu et al.(2008)]{Zhu08}       
       Zhu, Q.-F., Lacy, J. F., Jaffe, D. T., et al.  2008, \apjs, 177, 584
       
\bibitem[Zinnecker \& Yorke(2007)]{Zinnecker07} 
       Zinnecker, H., \& Yorke, H. W.  2007, Annu. Rev. Astron. Astrophys., 45, 481
\bibitem[Zinnecker(1982)]{Zinnecker82}
      Zinnecker, H.  1982, NYASA, 395, 226

\end{thebibliography}
\end{document}